\documentclass[useAMS,usenatbib]{mn2e}
\usepackage{graphicx}

\title[$\mu$ Eridani from MOST and from the ground]{$\bmu$ Eridani from 
MOST\thanks{Based in part on data from the MOST satellite, a Canadian Space Agency mission, jointly operated 
by Dynacon Inc., the University of Toronto Institute for Aerospace Studies and the University of British 
Columbia, with the assistance of the University of Vienna.} and from the ground: an orbit, the SPB 
component's fundamental parameters, and the SPB frequencies}

\author[M. Jerzykiewicz et al.]{M.~Jerzykiewicz,$^{1}$\thanks{E-mail: mjerz@astro.uni.wroc.pl} 
H.~Lehmann,$^{2}$ E.~Niemczura,$^{1}$ J.~Molenda-\.Zakowicz,$^{1}$\newauthor W.~Dymitrov,$^{3}$ 
M.~Fagas,$^{3}$ D.B.~Guenther,$^{4}$ M.~Hartmann,$^{2}$ M.~Hrudkov\'a,$^{2,5}$\newauthor K.~Kami\'nski,$^{3}$ 
A.F.J.~Moffat,$^{6}$ R.~Kuschnig,$^{7,8}$ G.~Leto,$^{9}$ J.M.~Matthews,$^{8}$\newauthor J.F.~Rowe,$^{10}$ 
S.M.~Ruci\'nski,$^{11}$ D.~Sasselov$^{12}$ and W.W.~Weiss$^{4}$\\
$^{1}$Instytut Astronomiczny Uniwersytetu Wroc{\l}awskiego, Kopernika 11, Pl-51-622 Wroc{\l}aw, Poland\\
$^{2}$Th\"uringer Landessternwarte Tautenburg, Sternwarte 5, D-07778 Tautenburg, Germany\\
$^{3}$Astronomical Observatory Institute, Faculty of Physics, A.~Mickiewicz University, S{\l}oneczna 36, 
Pl-60-286 Pozna\'n, Poland\\
$^{4}$Department of Astronomy and Physics, St.\ Mary's University Halifax, NS B3H 3C3, Canada\\
$^{5}$Isaac Newton Group of Telescopes, Apartado de Correos 321, E-387 00 Santa Cruz de la Palma, Canary 
Islands, Spain\\
$^{6}$D\'epartment de physique, Universit\'e de Montr\'eal, C.P.\ 6128, Succursale Centre-Ville, Montr\'eal, 
QC H3C 3J7, Canada\\
$^{7}$Institut f\"ur Astronomie, Universit\"at Wien, T\"urkenschanzstrasse 17, A-1180 Wien, Austria\\
$^{8}$Department of Physics and Astronomy, University of British Columbia, Vancouver, BC V6T1Z1, Canada\\
$^{9}$INAF-Osservatorio Astrofisico di Catania, via S.\ Sofia 78, I-95123 Catania, Italy\\
$^{10}$NASA Ames Research Center, Moffett Field, CA 94035, USA\\
$^{11}$Department of Astronomy and Astrophysics, University of Toronto, 50 St.\ George Street, Toronto, ON 
M5S 3H4, Canada\\
$^{12}$Harvard-Smithsonian Center for Astrophysics, 60 Garden Street, Cambridge, MA 02138, USA}

\input epsf
\usepackage{graphics}
\begin{document}

\date{Accepted ... Received ...; in original form ...}

\pagerange{\pageref{firstpage}--\pageref{lastpage}} \pubyear{2012}

\maketitle

\label{firstpage}

\begin{abstract} 

MOST time-series photometry of $\mu$ Eri, an SB1 eclipsing binary with a rapidly-rotating SPB primary, is 
reported and analyzed. The analysis yields a number of sinusoidal terms, mainly due to the intrinsic 
variation of the primary, and the eclipse light-curve. New radial-velocity observations are presented and 
used to compute parameters of a spectroscopic orbit. Frequency analysis of the radial-velocity residuals 
from the spectroscopic orbital solution fails to uncover periodic variations with amplitudes greater than 2 
km\,s$^{-1}$. A Rossiter-McLaughlin anomaly is detected from observations covering ingress.

>From archival photometric indices and the revised Hipparcos parallax we derive the primary's effective 
temperature, surface gravity, bolometric correction, and the luminosity. An analysis of a high 
signal-to-noise spectrogram yields the effective temperature and surface gravity in good agreement with the 
photometric values. From the same spectrogram, we determine the abundance of He, C, N, O, Ne, Mg, Al, Si, P, 
S, Cl, and Fe. 

The eclipse light-curve is solved by means of {\sc EBOP}. For a range of mass of the primary, a value of mean 
density, very nearly independent of assumed mass, is computed from the parameters of the system. Contrary to 
a recent report, this value is approximately equal to the mean density obtained from the star's effective 
temperature and luminosity. 

Despite limited frequency resolution of the MOST data, we were able to recover the closely-spaced SPB 
frequency quadruplet discovered from the ground in 2002-2004. The other two SPB terms seen from the ground 
were also recovered. Moreover, our analysis of the MOST data adds 15 low-amplitude SPB terms with 
frequencies ranging from 0.109 d$^{-1}$ to 2.786 d$^{-1}$. 

\end{abstract}


\section{Introduction}

The star $\mu$ Eri\,=\,HR\,1520\,=\,HD\,30211 (B5\,IV, $V={}$4.00 mag) was  discovered to be a single-lined 
spectroscopic binary by \citet{fl} at Yerkes Observatory. An orbit was derived by \citet{BvA} from radial 
velocities measured on 19 spectrograms taken in November and December 1956 with the 82-inch McDonald 
Observatory  telescope. The orbit was revised by \citet{h} who supplemented the McDonald observations with 
the Yerkes radial-velocities \citep*[published by][]{fbs}, Lick Observatory data of \citet{cm} and Dominion 
Astrophysical Observatory (DAO) observations of \citet{p}. Since these data span more than half a century, 
the value of the orbital period derived by \citet{h} is given with six-digit precision, viz., $7.35886$ d. 

The star was discovered to be variable in light by \citet{hsj} while being used as a comparison star in the 
2002-2003 multi-site photometric campaign devoted to $\nu$ Eri. A frequency analysis of the campaign data 
revealed a dominant frequency of 0.616 d$^{-1}$. Prewhitening with this frequency resulted in an amplitude 
spectrum with a strong $1/f$ component indicating complex variation. 

Using photometric indices and the {\em Hipparcos\/} parallax, \citet{hsj} determined the position of $\mu$ 
Eri in the HR diagram and compared it with evolutionary tracks. They found the star to lie close to a 6 
M$_{{\sun}}$ track, just before or shortly after the end of the main-sequence phase of evolution. Taking into 
account this result, the value of the dominant frequency, and the fact that the $u$ amplitude at this 
frequency is about a factor of two greater than the $v$ and $y$ amplitudes, \citet{hsj} concluded that $\mu$ 
Eri is probably an SPB star. 

\citet{hsj} have also considered the possibility that the dominant variation is caused by rotational 
modulation. Assuming the dominant frequency to be equal to the frequency of  rotation  of the star and using  
a value of the radius computed from the effective temperature and luminosity, they obtained $\sim$190 
km\,s$^{-1}$ for the equatorial velocity of rotation, $V_{\rm rot}$. This value and the published estimates 
of $V_{\rm rot} \sin i$, which range from 150 km\,s$^{-1}$ \citep{alg} to 190 km\,s$^{-1}$ \citep{bp}, make 
rotational modulation a viable hypothesis. However, \citet{hsj} judged the hypothesis of rotational 
modulation to be less satisfactory than pulsations because it does not explain the complexity of the light 
variation.   

The star was again used as the comparison star for $\nu$ Eri in the subsequent 2003-2004 multi-site 
photometric campaign \citep[][henceforth JHS05]{jhs}. The frequency of 0.616 d$^{-1}$ was confirmed and five 
further frequencies, ranging from 0.568 to 1.206 d$^{-1}$, were found. The multiperiodicity, the values of 
the frequencies and the decrease of the $uvy$ amplitudes with increasing wavelength  implied 
high-radial-order $g$ modes, strengthening the conclusion of \citet{hsj} that $\mu$ Eri is an SPB variable. 

An attempt at determining the angular degree and order, $\ell$ and $m$, of the six observed SPB terms was 
undertaken by \citet{ddp08}. Using a method of mode identification devised for SPB models in which the 
pulsation and rotation frequencies are comparable \citep{ddp07, dzdp07}, these authors obtained $\ell ={}$1 
and $m ={}$0 for the dominant mode. In addition, they estimated the expected amplitude of this mode's 
radial-velocity variation to be 5.9 km\,s$^{-1}$. For the remaining terms, there was some ambiguity 
in determining $\ell$ and $m$. Nevertheless, at least four terms detected in the star's light variation 
could be attributed to unstable $g$ modes of low angular degree provided that the star was assumed to be 
still in the main-sequence phase of its evolution. For these four terms, the predicted radial-velocity 
amplitudes ranged from 0.6 km\,s$^{-1}$ to 13.1 km\,s$^{-1}$. 

In addition to confirming the SPB classification, the analysis of the 2003-2004 campaign data showed $\mu$ 
Eri to be an eclipsing variable. As can be seen from figs.\ 3 and 5 of JHS05, the eclipse is a  
probably total transit, the secondary is fainter than the primary by several magnitudes, and the system is widely 
detached. The eclipse ephemeris, derived from the combined 2002-2003 and 2003-2004 campaign data, can be 
expressed as follows: 
\begin{equation} 
{\rm Min.\ light}={\rm HJD}\,2452574.04\,(4)+ 7.3806\,(10)\,E. 
\end{equation} 
Here, $E$ is the number of minima after the initial epoch given (which is that of the middle of the first 
eclipse, observed in 2002), and the numbers in parentheses are estimated standard deviations with the 
leading zeroes suppressed. 

The photometric orbital period in equation~(1) differs by more than 20 standard deviations from \citet*{h} 
period, mentioned in the first paragraph of this Introduction. \citet*{h} value is certainly less secure 
than the photometric value because it is based on a small number of observations, distributed unevenly over an 
interval of 53.2 years. In the periodogram of these observations, computed by \citet{j09}, the three highest 
peaks in the period range of interest, all of very nearly the same height, occurred at 7.35889, 7.38116, and 
7.38962 d. The first of these numbers corresponds to \citet*{h} value, while the second is equal to the 
photometric period to within half the standard deviation of the latter. Since both these numbers represent 
the radial-velocity data equally well, \citet{j09} concluded that the photometric period is the correct one. 

\citet{bs} observed $\mu$ Eri in February 2004 and August 2005 by means of the WIRE star tracker. After 
removing the SPB light-variation, they phased the observations with the orbital period of 7.381 d. \citet{bs} 
suggest that the eclipse depth is variable. 

The radial-velocity observations of \citet{BvA} and the photometric time-series data of \citet{hsj} and  
JHS05 were used by \citet{j09} to revise the spectroscopic orbital elements, solve the eclipse light-curve, 
and derive the primary's mean density. The mean density turned out to be smaller than the value obtained from 
the position in the HR diagram determined by \citet{hsj}, placing the star  beyond the TAMS. This conflicts 
not only with the star's photometric effective temperature, but also with the SPB classification because 
pulsational instability of $g$ modes disappears shortly after the minimum of $T_{\rm eff}$ in the 
main-sequence phase of evolution \citep{dmp}. Improving the orbital elements of $\mu$ Eri and thus obtaining 
an accurate value of the mean density of the primary component in order to find out how severe these 
conflicts are was the motivation for undertaking the present work. In addition, we expected to verify the 
frequencies of the six SPB terms found by JHS05 and discover further frequencies, given that the photometric 
amplitudes are too small to be detected from the ground. Finally, we planned to look for the SPB 
radial-velocity variations predicted by \citet{ddp08}. 

In the next section, we describe the MOST times-series photometry of $\mu$ Eri. In Section 3, we 
carry out a frequency analysis of these data and use the analysis results to remove the intrinsic component 
of the variation, thus bringing out the eclipse light-curve; in this section we also improve the eclipse 
ephemeris. In Section 4, we present new radial-velocity observations of $\mu$ Eri, some obtained 
contemporaneously with the MOST photometry, compute a spectroscopic orbit from them, and show that in the 
radial-velocity residuals from the orbital solution there are no periodic terms with an amplitude exceeding 
2 km\,s$^{-1}$. In Section 5, we solve the eclipse light-curve by means of the computer program {\sc EBOP} 
\citep{etz}. Section 6 deals with the Rossiter-McLaughlin effect. In Section 7, we determine the primary 
component's fundamental parameters from (1)~photometric indices and the revised {\em Hipparcos\/} parallax 
and (2) from high-resolution spectrograms; the spectrograms are also used to obtain the metal abundances by 
means of the method of spectrum synthesis. In Section 8, we derive the primary component's mean density and 
discuss the star's position in the HR diagram in relation to evolutionary tracks. Section 9 is devoted to 
assessing the SPB frequencies and amplitudes. The last section contains a summary of the results and 
suggestions for improvements. Short-period, small-amplitude variations found in the MOST observations of 
$\mu$ Eri are considered in the Appendix. 

\section{MOST time-series photometry}

\begin{figure} 
\includegraphics{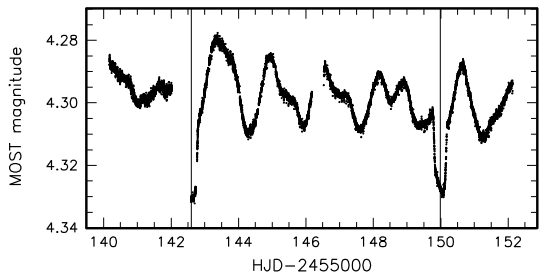} 
\caption{The MOST light-curve of $\mu$ Eri. The vertical lines indicate estimated epochs of mid-eclipse.}
\label{Fig01}
\end{figure}

MOST \citep{W+} observed $\mu$ Eri in 2009, from November 4  to 16. The light curve is shown in 
Fig.~\ref{Fig01}. The two wide gaps seen in the figure divide the light curve into three parts: (1) from the 
first point at HJD\,2455140.1633 to HJD\,2455142.0219, (2)~from~HJD\,2455142.5939 to HJD\,2455146.1774, and 
(3) from HJD\,2455146.5240 to the last point at HJD\,2455152.1273. Part 1 throughout and part 2 from the 
beginning to HJD\,2455145.1394 consist of 27 and 37 segments, respectively. The duration of most segments is 
equal to about 0.045~d, but there is one shorter segment in part 1 (0.0265 d), and two in part 2 (0.0098 d 
and 0.0356 d); the mean duration is equal to 0.0433 d. The mean duration of gaps between segments (not 
counting the gap between part 1 and part 2) amounts to 0.0259 d, and the mean interval between successive 
segments is equal to 101.4 min, the orbital period of the satellite. In part 2, from HJD\,2455145.1554 to the 
end and in part 3 throughout there are less than one-hundred gaps, all shorter than 0.012 d. The shortest 
sampling interval and the number of data-points per unit time differ between segments and parts. In all 
segments of part 1 and the first 28 segments of part 2, the shortest sampling interval is equal to 0.0004 d 
and there are about 111 data-points per hour; in the remaining nine segments and the non-segmented portion of 
part 2 and in part 3, the shortest sampling interval is equal to 0.0007 d and there are about 55 data-points 
per hour. 

The SPB variation time-scales are much longer than the duration of the data segments. It is therefore 
surprising that in many segments one can see short-term variations with ranges reaching 5 mmag. Short-term 
variations can be also seen in the non-segmented data. Further discussion of these variations we defer to 
the Appendix.

\section{The eclipse}

\subsection{Frequency analysis of the out-of-eclipse data}

In the light curve shown in Fig.~\ref{Fig01}, one can see the complex SPB variation and two successive 
eclipses. Unfortunately, in the first eclipse the ingress and the first half of the phase of totality are 
missing because of the first wide gap in the data. In addition, the eclipse is segmented (see Sect.\ 2), 
with the widest gap between segments (0.035 d) affecting the egress. We estimate the mid-eclipse epochs to 
be equal to HJD\,2455142.60$\,\pm\,$0.03 d and HJD\,2455149.98$\,\pm\,$0.02 d (see the vertical lines in 
Fig.~\ref{Fig01}). In order to obtain data suitable for analysing the SPB variation we modified the time series 
by omitting points falling within $\pm$0.03 orbital phase away from the mid-eclipse epochs. In the process, the 
number of data points was reduced from 15687 to 14765. 

\begin{figure} 
\includegraphics{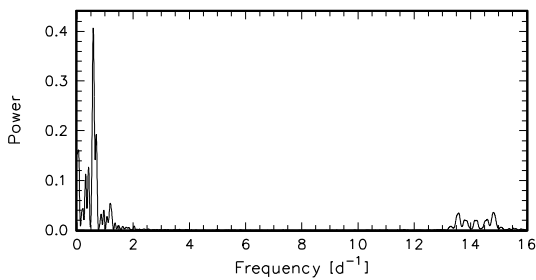} 
\caption{The power spectrum of the out-of-eclipse data.}
\label{Fig02}
\end{figure}

The power spectrum or periodogram of the modified data is shown in Fig.~\ref{Fig02}. Here and in what 
follows, by `power' we mean $1 - \sigma^2(f)/\sigma^2$, where $\sigma^2(f)$ is the variance of a 
least-squares fit of a sine curve of frequency $f$ to the data, and $\sigma^2$ is the variance of the same 
data. 

\begin{figure} 
\includegraphics{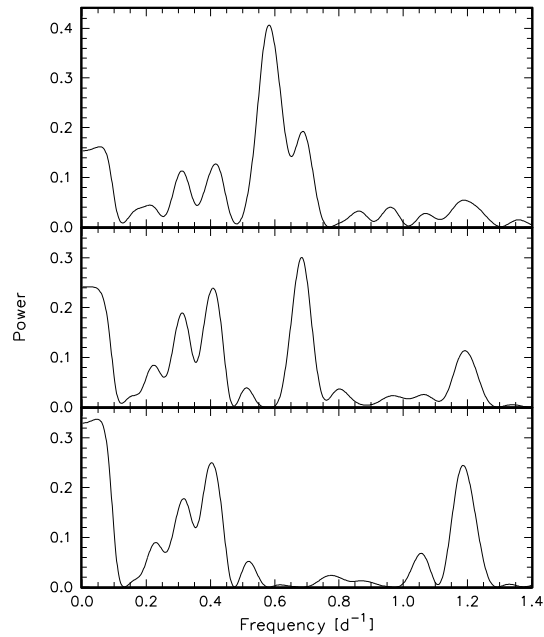} 
\caption{The power spectrum of the out-of-eclipse data (top), after prewhitening with the frequency $f_1 
={}$0.582 d$^{-1}$ (middle), and after prewhitening with $f_1$ and $f_2 ={}$0.684 d$^{-1}$ (bottom).}
\label{Fig03}
\end{figure}

As expected, the power in the periodogram in Fig.~\ref{Fig02} is concentrated in the frequency range $0 < f 
\le{}$1.4 d$^{-1}$, with a low-power alias around 14.2 d$^{-1}$, the orbital frequency of the satellite. 
This alias consists of the scaled-down pattern seen at low frequencies, shifted by 14.2 d$^{-1}$, and the 
mirror reflection about the $f ={}$14.2 d$^{-1}$ axis of the scaled-down pattern. The periodogram in the 
frequency range 0${}< f \le{}$1.4 d$^{-1}$ is displayed in the top panel of Fig.~\ref{Fig03}. The FWHM of 
the periodogram's highest lobe, centered at $f_1 ={}$0.582~d$^{-1}$, is very nearly equal to $1/T ={}$0.083 
d$^{-1}$, where $T ={}$12.0 d is the total time-span of the data. This shows that in our case the commonly 
used estimate of the frequency resolution, given by $1/T$, should be preferred over the more stringent value 
of $1.5/T$, advocated by \citet{ld}. 

In the middle panel of Fig.~\ref{Fig03} there is shown the periodogram of the out-of-eclipse data prewhitened 
with the frequency $f_1$. The highest lobe, centered at $f_2 ={}$0.684 d$^{-1}$, has already appeared as a 
just-resolved feature in the periodogram of the original data in the top panel of the figure. Finally, the 
bottom panel shows the periodogram obtained after prewhitening the data with the two frequencies, $f_1$ and 
$f_2$. The power at low frequencies seen in this panel indicates a variation on a time scale longer than $T$. 
The isolated lobe at right is centered at 1.186 d$^{-1}$. 

\begin{figure} 
\includegraphics[width=80mm]{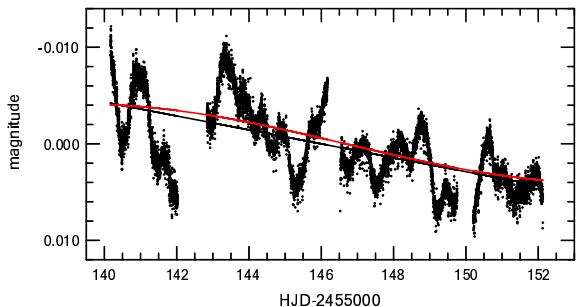} 
\caption{The out-of-eclipse magnitudes of $\mu$ Eri prewhitened with $f_1$, $f_2$, and 1.186 d$^{-1}$. The 
straight line (black) is a linear trend and the line running above it (red) is a segment of a sine-curve that 
has a period equal to a synodic month and maximum at the phase of Full Moon, fitted to the data together 
with the three frequencies. ({\em A colour version of this figure is available in the online journal.})}
\label{Fig04}
\end{figure}

Fig.~\ref{Fig04} shows the out-of-eclipse magnitudes prewhitened with $f_1$, $f_2$, and 1.186 d$^{-1}$. The 
long-term variation is clearly seen. A similar long-term variation was present in the MOST observations of 
$\delta$ Ceti. In the latter case it was shown by \citet{j07} to have been caused by inadequate compensation 
for background light in the MOST data reductions, so that the maximum of the variation occurred close to the 
phase of Full Moon, and the minimum, close to the phase of New Moon. The present case is similar. The upper 
line in Fig.~\ref{Fig04} represents a segment of a sine-curve having a period equal to a synodic month and 
its maximum fixed at the phase of Full Moon, fitted to the data together with the three frequencies. Also 
shown in the figure is a linear trend, fitted to the data together with the same frequencies. The near 
coincidence of the two lines proves our point. However, the assumption that the maximum of the long-term 
variation occurs {\em exactly\/} at the phase of Full Moon may be unwarranted because it is not known how 
much of the variation is caused by direct illumination from the Moon (which is a function not only of the lunar 
phase, but also of the angular distance of the Moon from the spacecraft), and how much is due to increased 
earth-shine on moonlit nights. In addition, assuming the variation to be strictly sinusoidal may be an 
oversimplification. Unfortunately, the total time-span of the data is shorter than 0.4 of a synodic month, 
so that the significance of these two factors cannot be evaluated. For the same reason, including the synodic 
frequency in the analysis may affect the low SPB frequencies. Therefore, we decided to account for the 
long-term variation by subtracting a linear trend. Thus, the observational equations to be fitted by the 
method of least squares to the out-of-eclipse magnitudes, $m_i$, were assumed to have the following form: 
\begin{equation} 
m_i = A_0 + B t_i + \sum_{j=1}^N A_j \sin (2 \pi f_j t_i + \Phi_j), 
\end{equation} 
where the second term accounts for the trend and $t_i ={}$HJD$-$2455146. 

The frequencies from $f_3$ to $f_N$ were then derived, one at a time, by determining the frequency of the 
highest peak in the periodogram computed after prewhitening with all frequencies already found and the trend. 
Adding the trend has affected the third frequency to be derived: it changed from 1.186 d$^{-1}$ to 1.193 
d$^{-1}$. An attempt to refine the frequencies by means of the method of non-linear least squares was 
abandoned  because (1) the ``refined'' frequencies differed grossly from their initial values, and (2) the 
solutions diverged for $N \ge{}$9. Parameters of equation (2) for $N ={}$31 were computed by means of the 
linear least squares fit with frequencies read off the periodograms. The standard deviation of the fit was 
equal to 0.64 mmag, the first term, $A_0 ={}$4297.843$\,\pm\,$0.008 mmag, and the second term, $B 
={}$0.6253$\,\pm\,$0.0026 mmag\,d$^{-1}$. The frequencies, $f_j$, the amplitudes, $A_j$, and the phases, 
$\Phi_j$, are listed in Table 1 in order of decreasing amplitude. The signal-to-noise ratios, $S/N$, given in 
the last column, were computed using $A_j$ as $S$, and as $N$, the mean amplitude over a 1 d$^{-1}$ interval 
around $f_j$ in the amplitude spectrum of the data prewhitened with all 31 frequencies. The decrease of $S/N$ 
with $A_j$ is not monotonic because $N$ is a function of $f_j$. The amplitudes, and therefore the order of 
terms in Table 1, depend on the number of terms taken into account. This is why the third frequency derived 
from the periodograms is $f_4$ in the table. The frequencies which differ by less than $1/T$ from a frequency 
of larger amplitude are indicated with an asterisk. We included these frequencies because omitting them would 
defy our purpose of removing as much out-of-eclipse variation as possible from the light-curve. By the same 
token, we included terms with $S/N <{}$4.0 and those with $f_j >{}$15~d$^{-1}$ although the former are 
insignificant according to the popular criterion of \citet{Bre93} while the latter are probably artefacts 
because they have frequencies that are either close to the whole multiples of the orbital period of the 
satellite or differ from the whole multiples by $\sim$1 d$^{-1}$. As shown in the Appendix, these frequencies 
are related to the short-term variations mentioned in Sect.\ 2. 

\begin{table}
 \centering
 \begin{minipage}{80mm}
  \caption{Frequencies, amplitudes and phases in the MOST out-of-eclipse magnitudes of $\mu$ Eri. The last 
column contains the signal-to-noise ratio.}
\begin{center}
  \begin{tabular}
{rcr@{\hspace{3pt}$\pm$\hspace{3pt}}lr@{\hspace{3pt}$\pm$\hspace{3pt}}lr} \\
  \hline
  $j$ & \multicolumn{1}{c}{$f_j$ [d$^{-1}$]} & \multicolumn{2}{c}{$A_j$ [mmag]} & 
\multicolumn{2}{c}{$\Phi_j$ [rad]} & \multicolumn{1}{r}{S/N} \\
 \hline
  1&\hspace{1pt} 0.582& 6.618 & 0.011& 1.550& 0.002     &254.5\\
  2&\hspace{1pt} 0.684& 4.567 & 0.011& 2.904& 0.002     &175.7\\
  3&\hspace{1pt} 0.387& 2.843 & 0.013& 4.789& 0.004     &109.3\\
  4&\hspace{1pt} 1.193& 2.523 & 0.009& 1.322& 0.004     & 84.1\\
  5&\hspace{1pt} 0.297& 1.812 & 0.012& 2.849& 0.008     & 69.7\\
  6&\hspace{1.6mm} 0.232*& 1.642 & 0.011& 1.134& 0.008  & 63.2\\
  7&\hspace{1pt} 0.835& 1.334 & 0.010& 4.901& 0.008     & 51.3\\
  8&\hspace{1.6mm} 1.254*& 1.200 & 0.008& 0.781& 0.007  & 40.0\\
  9&\hspace{1pt} 0.109& 1.177 & 0.011& 5.305& 0.008     & 45.3\\
 10&\hspace{1pt} 1.035& 1.058 & 0.009& 5.327& 0.009     & 37.8\\
 11&\hspace{1.6mm} 1.115*& 0.770 & 0.009& 2.744& 0.013  & 25.7\\
 12&\hspace{1pt} 1.825& 0.708 & 0.008& 3.724& 0.012     & 25.3\\
 13&\hspace{1.6mm} 0.526*& 0.556 & 0.010& 1.864& 0.021  & 21.4\\
 14&\hspace{1pt} 2.423& 0.446 & 0.008& 2.985& 0.018     & 17.2\\
 15&\hspace{1pt} 1.620& 0.333 & 0.009& 1.915& 0.025     & 11.1\\
 16&\hspace{1.6mm} 0.754*& 0.313 & 0.010& 1.773& 0.032  & 12.0\\
 17&\hspace{1pt} 2.682& 0.290 & 0.008& 2.399& 0.028     & 11.2\\
 18&\hspace{1pt} 2.199& 0.247 & 0.008& 2.547& 0.033     &  8.8\\
 19&\hspace{1pt} 1.722& 0.227 & 0.008& 3.759& 0.036     &  8.1\\
 20&\hspace{1pt} 2.786& 0.164 & 0.008& 2.251& 0.048     &  6.3\\
 21& 71.960& 0.152 & 0.008& 4.714& 0.052                & 11.3\\
 22&\hspace{1pt} 3.780& 0.152 & 0.008& 0.523& 0.050     &  3.5\\
 23& 28.432& 0.142 & 0.007& 0.025& 0.054                &  7.5\\
 24&\hspace{1pt} 2.301& 0.142 & 0.008& 3.833& 0.058     &  5.5\\
 25& 15.190& 0.137 & 0.008& 0.314& 0.057                &  5.8\\
 26& 57.766& 0.135 & 0.008& 1.082& 0.058                &  9.6\\
 27&\hspace{1pt} 4.850& 0.126 & 0.008& 1.429& 0.059     &  3.5\\
 28& 42.524& 0.124 & 0.008& 0.391& 0.061                &  6.2\\
 29& 55.764& 0.110 & 0.008& 5.540& 0.068                &  7.3\\
 30& 86.159& 0.110 & 0.008& 1.583& 0.070                &  7.3\\
 31&\hspace{1pt} 5.499& 0.101 & 0.007& 5.695& 0.074     &  3.0\\
\hline
\end{tabular}
\end{center}
\end{minipage}
\end{table}

\begin{figure} 
\includegraphics{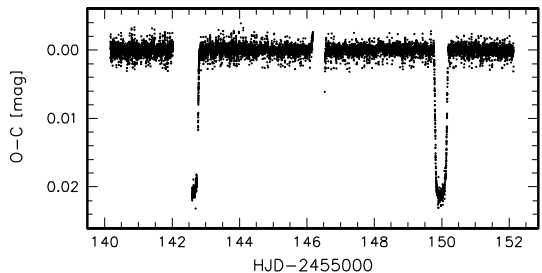} 
\caption{The residuals, $O-C$, from a least-squares fit of equation (2) with $N ={}$31 to the 
out-of-eclipse data.}
\label{Fig05}
\end{figure}

\subsection{The eclipse light-curve and the corrected ephemeris}

The residuals, $O-C$, where $O$ is the observed magnitude and $C$ is the magnitude computed from a 
least-squares fit of equation (2) with $N ={}$31 to the out-of-eclipse data, are plotted in 
Fig.~\ref{Fig05}. From the residuals, the mid-eclipse epoch of the second eclipse could be derived much more 
accurately than was possible with the original data (see the first paragraph of Sect.\ 3.1). The result 
is HJD\,2455149.973$\,\pm\,$0.002. Using this number and equation (1) we arrive at the following corrected 
ephemeris:
\begin{equation} 
{\rm Min.\ light}={\rm HJD}\,2455149.973\,(2)+ 7.38090\,(12)\,E, 
\end{equation} 
where the notation is the same as before, except that the minima are now counted from the new, more accurate 
initial epoch. Note that the new value of the orbital period differs from the old value by about 0.3 of the 
standard deviation of the latter. 

\section{The spectroscopic orbit}

\subsection{The radial-velocity data}

On 24 nights between 2 October 2009 and 8 February 2010, HL, MHa, and MHr obtained 105 spectrograms of $\mu$ 
Eri with the coud\'e echelle spectrograph at the 2-m Alfred Jensch telescope of the Th\"uringer 
Landessternwarte Tautenburg (TLS). The spectral resolving power was 63,000. The spectrograms were 
reduced using standard MIDAS packages. Reductions included bias and stray light subtraction, filtering of 
cosmic rays, flat fielding, optimum extraction of echelle orders, wavelength calibration using a ThAr lamp, 
correction for instrumental shifts using a large number of O$_2$ telluric lines, normalization to local 
continuum, and weighted merging of orders. 

On eight nights, 4 to 14 November 2009, JM-\.Z obtained 43 spectrograms of $\mu$ Eri with the 91-cm 
telescope and the fiber-fed echelle spectrograph FRESCO of the M.G.\ Fracastoro station of the Osservatorio 
Astrofisico di Catania (OAC). The spectral resolving power was 21,000. The spectrograms were 
reduced with IRAF\footnote{IRAF is distributed by the National Optical Astronomy Observatory, which is 
operated by the Association of Universities for Research in Astronomy, Inc.}. 

Finally, on eight nights between 29 September 2009 and 22 January 2010, WD obtained 64 spectrograms of the 
star with the Pozna\'n Spectroscopic Telescope\footnote{www.astro.amu.edu.pl/PST/} (PST). The reductions 
were carried out by MF and KK by means of the Spectroscopy Reduction 
Package\footnote{http://www.astro.amu.edu.pl/$\sim$chrisk/MOJA/pmwiki.php?n= SRP.SRP}.

Radial velocities were determined from the spectrograms by EN by means of cross-correlation with a template 
spectrum in the logarithm of wavelength. A synthetic spectrum, computed assuming the photometric $T_{\rm 
eff}$ and $\log g$ (see Sect.\ 7.1), a microturbulence velocity of 4.0 km\,s$^{-1}$, $V_{\rm rot} \sin i 
={}$136 km\,s$^{-1}$, solar helium abundance, and the EN metal abundances from Table 6, served as the 
template. The wavelength intervals used in the cross-correlations included He\,I lines and lines of the 
elements listed in Table 6, had high signal-to-noise ratio (S/N), and were not affected by telluric lines. 
The S/N were determined with the IRAF task {\sf splot} (function "m") from the continuum in the 5100 \AA\ to 
5800 \AA\ range which contains only very faint lines. The S/N (per pixel) of the spectrograms ranged from 70 
to 270 for TLS, from 60 to 110 for OAC, and from 40 to 120 for PST; the overall mean S/N values were  
198$\,\pm\,$3, 93$\,\pm\,$2, and 78$\,\pm\,$2 for TLS, OAC, and PST, respectively. 

\begin{table}
 \centering
 \begin{minipage}{80mm}
  \caption{Nightly mean radial-velocities of $\mu$ Eri.}
\begin{center}
  \begin{tabular}
{lccrc} \\
  \hline
\multicolumn{1}{c}{$\langle$HJD$\rangle$} & \multicolumn{1}{c}{$\langle$RV$\rangle$} & \multicolumn{1}{c}
{$\langle$S/N$\rangle$} & \multicolumn{1}{c}{N} & \multicolumn{1}{c}{Obs}\\
\multicolumn{1}{c}{ } & \multicolumn{1}{c}{[km\,s$^{-1}$]} & \multicolumn{1}{c}{ } & \multicolumn{1}{c}{ } & 
\multicolumn{1}{c}{ }\\
 \hline
 2455104.6491&\hspace{7pt}28.35&\hspace{3pt}87.7 & 3 & PST  \\
 2455106.5913&\hspace{4pt}$-$8.02&    191.5 & 2 & TLS  \\
 2455107.6194&\hspace{7pt}13.82&    179.4 & 2 & TLS  \\
 2455113.6150&\hspace{4pt}$-$7.24&    210.0 & 1 & TLS  \\
 2455135.5878&\hspace{4pt}$-$4.29&    200.5 & 2 & TLS  \\
 2455136.6210&\hspace{4pt}$-$4.76&    210.0 & 1 & TLS  \\
 2455140.5652&\hspace{7pt}36.22&\hspace{3pt}91.7 & 4 & OAC  \\
 2455141.4761&\hspace{7pt}30.84&    110.5 & 4 & PST  \\
 2455141.5661&\hspace{7pt}30.12&    226.0 & 2 & TLS  \\
 2455142.5027&\hspace{7pt}13.82&    180.0 & 1 & TLS  \\
 2455143.5048& $-$11.36        &    150.0 & 1 & TLS  \\
 2455143.5660& $-$10.49        &\hspace{3pt}94.8 & 5 & OAC  \\
 2455144.4553&\hspace{8pt} 4.17&    110.0 & 1 & OAC  \\
 2455146.5152&\hspace{7pt}32.59&\hspace{3pt}87.6 & 2 & OAC  \\
 2455147.5578&\hspace{7pt}36.65&    216.0 & 2 & TLS  \\
 2455147.5767&\hspace{7pt}37.65&\hspace{3pt}96.4 & 7 & OAC  \\
 2455148.5484&\hspace{7pt}35.66&\hspace{3pt}92.5 & 7 & OAC  \\
 2455149.5758&\hspace{7pt}21.62&\hspace{3pt}88.3 & 8 & OAC  \\
 2455150.5478&\hspace{4pt}$-$5.55&    100.7 & 9 & OAC  \\
 2455151.4388&\hspace{4pt}$-$2.40&\hspace{3pt}75.3 & 4 & PST  \\
 2455155.4978&\hspace{7pt}37.30&\hspace{3pt}72.6 &21 & PST  \\
 2455155.5034&\hspace{7pt}37.10&    215.1 & 2 & TLS  \\
 2455156.5190&\hspace{7pt}30.64&\hspace{3pt}67.3 & 8 & PST  \\
 2455157.4154&\hspace{8pt} 9.53&\hspace{3pt}69.0 & 3 & PST  \\
 2455158.5251&\hspace{4pt}$-$9.05&  240.0 & 1 & TLS  \\
 2455161.4549&\hspace{7pt}37.23&\hspace{3pt}87.7 &14 & PST  \\
 2455162.4466&\hspace{7pt}32.40&    180.0 & 1 & TLS  \\
 2455163.4780&\hspace{7pt}32.42&    170.0 & 1 & TLS  \\
 2455168.4333&\hspace{7pt}31.12&    190.0 & 1 & TLS  \\
 2455170.4825&\hspace{7pt}37.45&    255.1 & 2 & TLS  \\
 2455175.3956&\hspace{7pt}29.30&    230.0 & 1 & TLS  \\
 2455192.4014&\hspace{7pt}36.54&    215.1 & 4 & TLS  \\
 2455194.3866&\hspace{8pt} 2.94&    230.0 & 1 & TLS  \\
 2455199.3759&\hspace{7pt}37.07&    179.8 &15 & TLS  \\
 2455201.4046&\hspace{7pt}21.68&    225.1 & 4 & TLS  \\
 2455202.3392& $-$12.10        &    191.4 &34 & TLS  \\
 2455219.3368&\hspace{7pt}25.07&\hspace{3pt}76.0 & 7 & PST  \\
 2455228.3739&\hspace{7pt}39.52&    260.0 & 1 & TLS  \\
 2455231.2637&\hspace{8pt} 3.63&    270.0 & 1 & TLS  \\
 2455236.2367&\hspace{7pt}34.70&    200.0 & 1 & TLS  \\
\hline
\end{tabular}
\end{center}
\end{minipage}
\end{table}

\subsection{The spectroscopic elements}

\begin{figure} 
\includegraphics{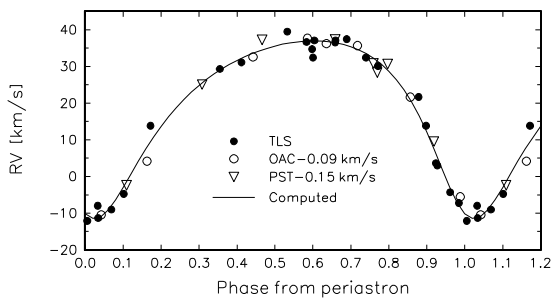} 
\caption{The TLS, OAC, and PST radial-velocities from Table~2, with the systematic corrections in the cases 
OAC and PST as indicated, plotted as a function of orbital phase. Also shown is the radial-velocity curve 
(line), computed from the spectroscopic elements of Table 3.}
\label{Fig06}
\end{figure}

The number of spectrograms taken on a night ranged from one to 34. Using the individual radial-velocities of 
the preceding section for deriving a spectroscopic orbit would give excessive weight to the few nights with a  
large number of observations. To avoid this, we took weighted means of the individual velocities on nights 
with two or more observations using the S/N of the individual spectrograms as weights; the epochs of 
observations and S/N were likewise averaged. In Table 2 are given the single ($N ={}$1) or the nightly 
weighted mean ($N \ge{}$2) epochs of observations, $\langle$HJD$\rangle$, the radial velocities, 
$\langle$RV$\rangle$, and the signal-to-noise ratios, $\langle$S/N$\rangle$. In the case of JD\,2455201, the 
means were computed from the four points unaffected by the Rossiter-McLaughlin effect (see below, Sect.\ 6). 
Using the $\langle$HJD$\rangle$, $\langle$RV$\rangle$, and $\langle$S/N$\rangle$ from Table 2, we computed a 
spectroscopic orbit by means of the non-linear least squares method of \citet{s}. We applied weights 
proportional to $\langle$S/N$\rangle$. The orbital period was assumed to be equal to the photometric period  
given in equation (3). An examination of the residuals from the solution revealed small systematic shifts 
between the TLS, OAC, and PST radial-velocities. Corrections to the OAC and PST radial-velocities which 
minimized the standard deviation of the solution turned out to be  $-$0.09$\,\pm\,$0.73 km\,s$^{-1}$ 
and $-$0.15$\,\pm\,$0.71 km\,s$^{-1}$ for OAC and PST, respectively. The elements of the orbit, computed 
with these corrections taken into account, are listed in Table 3. Figure~\ref{Fig06} shows the TLS 
radial-velocities and the corrected OAC and PST radial-velocities, plotted as a function of phase of the 
orbital period. The solid line is the radial-velocity curve computed from the elements listed in Table 3.

\begin{table}
 \centering
 \begin{minipage}{80mm}
  \caption{Orbital elements of $\mu$ Eri computed from the radial-velocities of Table 2, with the systematic 
corrections added to the OAC and PST data (see text and Fig.~\ref{Fig06}).}
\begin{center}
  \begin{tabular}
{ll}
  \hline
$P_{\rm orb}$ (assumed)&\hspace*{25pt}7.38090 d \\
$T$&HJD\,2455143.254\,$\pm$\,0.067 d\\
$\omega$&\hspace*{11.5pt}160.5$\degr\,\pm$\,4.5$\degr$\\
$e$&\hspace*{16pt}0.344\,$\pm$\,0.021\\
$\gamma$&\hspace*{16pt}20.58\,$\pm$\,0.43 km\,s$^{-1}$\\
$K_1$&\hspace*{16pt}24.24\,$\pm$\,0.53 km\,s$^{-1}$\\
$a_1\sin i$&\hspace*{16pt}3.320\,$\pm$\,0.077 R$_{\sun}$\\
$f(M)$&\hspace*{7pt}0.00902\,$\pm$\,0.00063 M$_{\sun}$\\
  \hline
\end{tabular}
\end{center}
\end{minipage}
\end{table}

\subsection{Looking for intrinsic radial-velocity variations}

In an attempt to detect intrinsic radial-velocity variations, we examined the residuals from the orbital 
velocity curve of the 40 nightly mean radial-velocities used to derive it. A periodogram of these data, 
computed with weights proportional to $\langle$S/N$\rangle$ (see Table 2), showed the highest peak at $f 
={}$0.812 d$^{-1}$. Interestingly, this frequency is close to the JHS05 frequency ($f'_3$ in their notation) 
for which \citet{ddp08} predict the largest velocity amplitude (see the Introduction). However, the 
predicted amplitude of $f'_3$ was equal to 13.1 km\,s$^{-1}$, while the amplitude of $f$ is equal to only 
1.9$\,\pm\,$0.4 km\,s$^{-1}$. In addition, a periodogram of the residuals from the orbital velocity curve of 
the individual radial-velocity measurements (without those affected by the Rossiter-McLaughlin effect) 
showed the highest peak with the same amplitude but at a frequency of 0.626 d$^{-1}$, unrelated to $f$. We 
conclude that both these frequencies, $f$ and 0.626 d$^{-1}$, are spurious, and that in the radial-velocity 
residuals there are no periodic terms with an amplitude exceeding 2 km\,s$^{-1}$. 

Unlike the PST and TLS observations, which fall mainly outside the interval covered by the MOST observations 
of $\mu$ Eri, all 43 OAC radial-velocities were obtained contemporaneously with the MOST photometry. Taking 
advantage of this, we have looked for a correlation between the OAC radial-velocity residuals and the 
trend-corrected MOST magnitudes. No correlation was found. Thus, the largest deviations from the orbital 
velocity curve, amounting to about $\pm\,$4.7 km\,s$^{-1}$ (see Fig.~\ref{Fig06}), are probably instrumental. 

\section{The eclipse light-curve solution}

For computing the parameters of $\mu$ Eri from the eclipse light-curve we used \citet*{etz} computer program 
{\sc EBOP}, which is based on the Nelson-Davis-Etzel model \citep{nd,pe}. The program is well suited for 
dealing with detached systems such as the present one. The spectroscopic parameters $\omega$ and $e$, needed 
to run the program, were taken from Table 3. The primary's limb-darkening coefficient we computed by 
convolving the MOST KG filter transmission \citep{W+} with the monochromatic limb-darkening coefficients for 
$T_{{\rm eff}} ={}$15\,670~K and $\log g ={}$3.5 (see Section 7.1) and [M/H] = 0 (Section 7.2). The 
monochromatic limb-darkening coefficients were interpolated in the tables of Walter Van 
Hamme\footnote{http://www2.fiu.edu/$\sim$vanhamme/limdark.htm, see also \citet{VH}}. For the primary, the 
result was 0.308. For the secondary, assuming the effective temperature to scale as the square root of the 
radius and taking the ratio of the radii to be equal to 0.135 (see Table 4), we obtained $T_{\rm eff} 
={}$5\,758~K. Using this value, we found the central surface brightness of the secondary in units of that of 
the primary (which {\sc EBOP} uses as a fundamental parameter) to be equal to 0.018. Assuming further the 
solar value of 4.4 for $\log g$ and solar metallicity, and using the same tables of monochromatic 
limb-darkening coefficients as above, we computed the secondary's limb-darkening coefficient to be equal to 
0.628. The remaining parameters we set to their default {\sc EBOP} values. The unknowns to be solved for 
were the orbital inclination, $i$, the ratio of the radii, $k = R_{\rm s}/R_{\rm p}$, and the radius of the 
primary, $r_{\rm p} = R_{\rm p}/a$, in units of the semi-major axis of the relative orbit, $a$. Initial 
values of the unknowns were derived by trial and error. Tests showed that the results were insensitive to 
the initial values. 

\begin{figure} 
\includegraphics{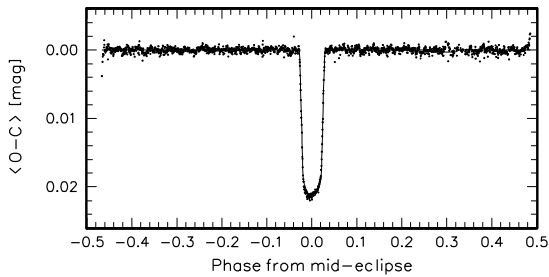} 
\caption {Normal points computed in adjacent intervals of 0.0005 orbital phase using the $O-C$ residuals 
shown in Fig.~\ref{Fig05}. The solid line is the synthetic light-curve, computed with the parameters (4) of 
Table 4.} 
\label{Fig07} 
\end{figure}

\begin{figure} 
\includegraphics{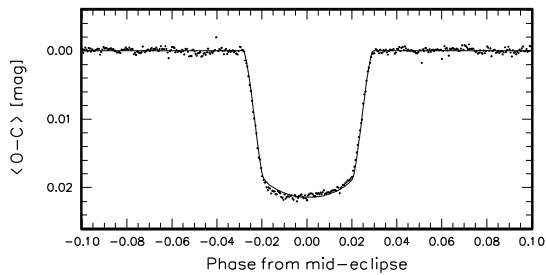} 
\caption{The same as Fig.~\ref{Fig07}, but for a limited interval of orbital phase.} 
\label{Fig08} 
\end{figure}

The results depend slightly on the light-curve data used. In Table 4 are shown the parameters obtained 
using (1)~the $O-C$ residuals from a least-squares fit of equation (2) with $N ={}$20 to the out-of-eclipse 
data, i.e., without terms $f_j >{}$15~d$^{-1}$, (2) normal points computed from the $O-C$ residuals in 
adjacent intervals of 0.0005 orbital phase; (3) and (4): the same as (1) and (2), respectively, but with $N 
={}$31. A synthetic light-curve computed with the parameters derived from data (4) (see Table 4) is compared 
with the data in Fig.~\ref{Fig07}; for orbital phases from $-$0.1 to 0.1, the synthetic light-curve is 
compared with the data in Fig.~\ref{Fig08}. 

\begin{table}
 \centering
 \begin{minipage}{87mm}
  \caption{The parameters of $\mu$ Eri derived by means of {\sc EBOP}. Data (1) ... (4) are explained in the 
text. SE is the {\sc EBOB} standard error of one observation in mmags.}
\begin{center}
\begin{tabular}{ccrcl} \\
\hline
\multicolumn{1}{c}{Data} & \multicolumn{1}{c}{SE} & \multicolumn{1}{c}{$i$} &
\multicolumn{1}{c}{$k$} & \multicolumn{1}{c}{$r_{\rm p}$}\\
\hline
(1)&0.71&82.06$\pm$0.07& 0.13507$\pm$0.00011&0.2094$\pm$0.0005\\
(2)&0.48&82.11$\pm$0.12& 0.13524$\pm$0.00021&0.2092$\pm$0.0008\\
(3)&0.64&82.42$\pm$0.07& 0.13502$\pm$0.00010&0.2072$\pm$0.0004\\
(4)&0.41&82.44$\pm$0.11& 0.13539$\pm$0.00018&0.2071$\pm$0.0007\\
  \hline
\end{tabular}
\end{center}
\end{minipage}
\end{table}

As can be seen from Fig.~\ref{Fig08}, the fit of the computed light-curve (line) to the observations 
(points) is not perfect at the bottom of the eclipse: before mid-eclipse the points fall below the lines, 
i.e., the system is fainter than computed, while after mid-eclipse the reverse is true. This looks like a 
signature of the rotational Doppler beaming, i.e., photometric Rossiter-McLaughlin effect \citep{Gr12,Sh+12}. 
However, from equation (3) of \citet{Sh+12} it follows that in the present case the amplitude of the effect 
would be an order of magnitude smaller than the observed deviation. 

The depth of the secondary eclipse predicted by the EBOP solution is 0.3 mmag, much smaller than the noise 
seen in Fig.~\ref{Fig07}. In the infrared, the depth of the secondary eclipse will be greater because the 
contribution of the primary to the total light of the system would decrease with increasing wavelength. The 
predicted MOST magnitude difference between components is 8.8 mag. Assuming that this number represents the 
magnitude difference in the $V$ band and using the standard $V-J$ and $V-K$ colour indices, one gets 2~mmag 
and 3 mmag as the depths of the secondary eclipse in J and K, respectively. Thus, detecting the secondary 
eclipse of $\mu$ Eri would be a task for a future infrared satellite. 

\section{The Rossiter-McLaughlin effect}

\begin{figure} 
\includegraphics{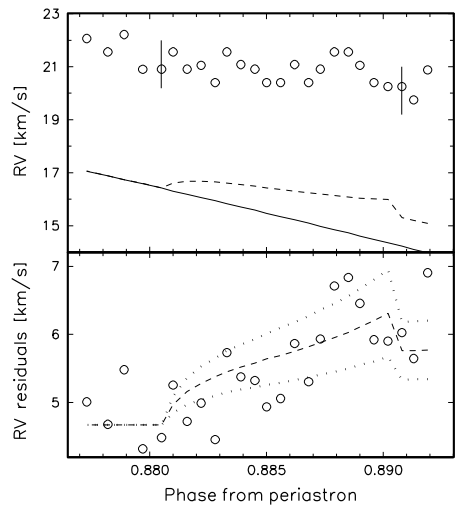} 
\caption{Upper panel: individual radial-velocity measurements on JD\,2455201 (open circles) plotted as a 
function of orbital phase. The short vertical lines indicate the epochs of first and second contact. 
Also shown are the radial-velocity curve, computed from the spectroscopic elements of Table 3 (solid line), 
and the radial-velocity curve plus the anomaly due to the Rossiter-McLaughlin effect, computed as explained 
in the text for $V_{\rm rot} \sin i ={}$115 km\,s$^{-1}$ (dashed line). Lower panel: radial-velocity 
residuals (open circles) and the Rossiter-McLaughlin anomaly for $V_{\rm rot} \sin i ={}$115 km\,s$^{-1}$, 
shifted upwards by 4.67 km\,s$^{-1}$ (dashed line), bracketed by those for $V_{\rm rot} \sin i ={}$70 
km\,s$^{-1}$ and $V_{\rm rot} \sin i ={}$160 km\,s$^{-1}$ (the lower and upper dotted line, respectively).}
\label{Fig09}
\end{figure}

On JD\,2455201, there were 25 measurements, the first one at HJD\,2455201.3958, and the last at 
HJD\,2455201.5034. According to equation (3), the corresponding orbital phases (reckoned from the epoch of 
mid-eclipse) are equal to $-$0.033 and $-$0.018. Thus, the measurements start shortly before the ingress 
begins, and last to just after it ends. This is illustrated in the upper panel of Fig.~\ref{Fig09}, where 
the epochs of first and second contact, predicted by the eclipse solution (4) of Table 4, are indicated 
by short vertical lines. Also shown in the figure is a fragment of the orbital velocity curve, computed from 
the spectroscopic elements of Table 3 (solid line in the upper panel), and the deviations from this curve 
due to the Rossiter-McLaughlin effect (dashed line), computed assuming (1) the direction of rotation 
concordant with the direction of orbital motion, and (2) $V_{\rm rot} \sin i$ of the primary component equal 
to 115 km\,s$^{-1}$, a value that gave the best least-squares fit to the data. The lower panel 
of Fig.~\ref{Fig09} shows the fit to the residuals, bracketed by the Rossiter-McLaughlin deviations for 
$V_{\rm rot} \sin i$ equal to 70 km\,s$^{-1}$ and 160 km\,s$^{-1}$ (the lower and upper dotted line, 
respectively). In computing the deviations we used formulae developed by \citet{p38}. These formulae assume 
the axis of rotation of the star to be perpendicular to the orbital plane, neglect limb darkening, and are 
valid for circular orbits. The last assumption was moderated to first order in the eccentricity by means of 
the prescription given by \citet{r12}. Taking into account the effect of limb darkening would reduce the 
computed deviations (see \citealt*{Oh05}), so that the best fitting $V_{\rm rot} \sin i$ would be greater by 
a factor of $\sim$1.5. We conclude that the increase of the radial-velocity residuals from about 4.5 
km\,s$^{-1}$ to about 6 km\,s$^{-1}$, seen in the lower panel of Fig.~\ref{Fig09}, is consistent with 
$V_{\rm rot} \sin i$ of about 170$\,\pm\,$45 km\,s$^{-1}$, but we postpone a more detailed discussion of the 
Rossiter-McLaughlin effect in $\mu$ Eri until more accurate data become available. Note, however, that 
although the value of $V_{\rm rot} \sin i$ is not well constrained at this stage, the direction of rotation 
is determined beyond doubt. 

On JD\,2455201, the mean residual from the orbital solution amounted to 5.49$\,\pm\,$0.15 km\,s$^{-1}$. 
Subtracting the Rossiter-McLaughlin deviations (dashed line in Fig.~\ref{Fig09}) reduces this value to 
4.67$\,\pm\,$0.10 km\,s$^{-1}$. The residual is probably instrumental (see Sect.\ 4.3).

\section{Stellar parameters}

\subsection{From photometric indices and the revised {\em Hipparcos\/} parallax}

The colour excess of $\mu$ Eri is small but not negligible. Using the star's $B-V$ and $U-B$ colour indices 
from \citet{m91} and \citet*{j66} standard two-colour relation for main-sequence stars, we get $E(B-V) 
={}$0.018$\,\pm\,$0.006 mag. From the Str\"omgren $uvby$ indices \citep{hm} and \citet*{c78} $(b-y)_0$ - 
$c_0$ relation, we have $E(b-y) ={}$0.013$\,\pm\,$0.004 mag or $E(B-V) ={}$0.018$\,\pm\,$0.005 mag. The 
dereddened colours and the $\beta$ index \citep{hm} then yield the effective temperature, $T_{\rm eff}$, the 
bolometric correction, BC, and the logarithmic surface gravity, $\log g$. The results are listed in Table 5, 
in columns three, four, and five, respectively. The BC values obtained using the calibrations of \citet{b94} 
and \citet{f96} were adjusted slightly (by 0.03 mag and 0.01 mag, respectively) in order to make them 
consistent with BC$_{\sun} ={}-$0.07 mag, the zero point adopted by \citet{cod}. As can be seen from Table 5, 
\citet*{f96} $T_{\rm eff}$ and BC are much lower than the other ones. The former values were obtained by 
interpolation from this author's table 3. Using the polynomial fits provided by \citet{f96} and corrected by 
\citet{torr} results in $T_{\rm eff}$ and BC virtually identical with those given in Table 5. Omitting 
\citet*{f96} values, we get 15\,670 K and $-$1.40 mag for the mean $T_{\rm eff}$ and BC, respectively. Our 
value of $T_{\rm eff}$ is identical to those derived by \citet{hsj} and \citet{ddp08} from Geneva 
photometry. However, while \citet{hsj} estimated the standard deviation of their value to be equal to 100 K, 
we believe that the standard deviation of $T_{\rm eff}$ cannot be smaller than the uncertainty of 
temperature calibrations, i.e., about 3 percent \citep{n93,j94} or 470 K for the $T_{\rm eff}$ in question. 
\citet{ddp08} give a somewhat more optimistic value of 365 K. 

Taking into account the uncertainty of calibrations, we estimate the standard deviation of the bolometric 
correction to be equal to 0.08 mag. Finally, the mean value of $\log g$ is equal to 3.5 dex (see Table 5), 
with an estimated standard deviation of 0.15 dex. 

\begin{table}
 \centering
 \begin{minipage}{90mm}
  \caption{Fundamental parameters of $\mu$ Eri obtained from the photometric indices listed in column one, 
using the calibration referenced in column two.}
\begin{center}
\begin{tabular}{llccc} \\
\hline
\multicolumn{1}{c}{Data} & \multicolumn{1}{c}{Calibration} & \multicolumn{1}{c}{$T_{\rm eff}$} &
\multicolumn{1}{c}{BC} & \multicolumn{1}{c}{$\log g$}\\
\hline
$(B-V)_0$&\citet{cod}&15\,700&$-$1.47&--\\
$(B-V)_0$&\citet{f96}&14\,610&$-$1.17&--\\
$c_0$    &\citet{ds77}&15\,670&$-$1.46&--\\
$c_0$, $\beta$&UVBYBETA\footnote{A FORTRAN program based on the grid published by \citet{md85}. 
Written in 1985 by T.T.\ Moon of the University London and modified in 1992 and 1997 by R.\ Napiwotzki 
of Universitaet Kiel.}&15740&--&3.44\\
$c_0$, $\beta$&\citet{b94}&15\,580&$-$1.28&3.52\\
  \hline
\end{tabular}
\end{center}
\end{minipage}
\end{table}

>From the revised {\em Hipparcos\/} parallax \citep{vL}, the $V$ magnitude from \citet{m91}, $E(B-V)$ from the 
first paragraph of this section, and assuming $R_V ={}$3.1, we get $M_V ={}-$2.06$\,\pm\,$0.07 mag, $M_{\rm 
bol} ={}-$3.46$\,\pm\,$0.10 mag, and $\log L/{\rm L}_{\sun} ={}$3.280$\,\pm\,$0.040. The standard deviations 
of $M_{\rm bol}$ and $\log L/$L$_{\sun}$ are determined by the standard deviation of BC and the standard 
deviation of $M_V$ (in that order). In computing $\log L/$L$_{\sun}$, we assumed M$_{{\rm bol}{\sun}} ={}$4.74 
mag, a value consistent with BC$_{\sun} =-{}$0.07 mag we adopted and $V_{\sun} ={}-$26.76 mag \citep{torr}.

\subsection{From TLS spectra}

A Fourier-transform based program KOREL \citep[][$\!$b]{Ha06a} was used to decompose the stellar spectrum and 
the telluric contributions with the orbital elements obtained in Section 4.2 adopted as starting values and 
the orbital period fixed at the photometric value. The result was a high-S/N spectrum of $\mu$ Eri, free of 
telluric contributions, representing a mean spectrum, averaged over all RV-shifted spectra of the TLS 
time-series. The analysis of this mean spectrum was carried out independently by HL and EN. HL used the 
program SynthV \citep{Ts96} to compute a grid of synthetic spectra from LTE model atmospheres calculated with 
the program LLmodels \citep{Sh+04}. For a detailed description of the method see \citet{Le+11}. For the 
analysis, the full wavelength range 4720~\AA\ to 7300~\AA\ was used. The atomic line parameters were adopted 
from the Vienna Atomic Line Database (VALD)\footnote{http://ams.astro.univie.ac.at/$\sim$vald, see also 
\citet{kup99}}. HL has derived the following parameters: $T_{\rm eff} ={}$ 15\,590$\,\pm\,$120 K, $\log g 
={}$3.55$\,\pm\,$0.04, the microturbulence velocity, $\xi ={}$0.0$\,\pm\,$0.4 km\,s$^{-1}$, and $V_{\rm rot} 
\sin i ={}$130$\,\pm\,$3 km\,s$^{-1}$. For the He abundance, HL has obtained 1.3$\,\pm\,$0.1 times the solar 
value. The errors are 1$\sigma$ errors from $\chi^2$ statistics as described in \citet{Le+11}. They represent 
a measure of the quality of the fit but do not account for imperfections of the model. Small corrections to 
the local continua of the observed spectra have been applied by comparing them to the synthetic spectra in the 
process of parameter determination. Possible correlations between the parameters are included by deriving the 
optimum values and their errors from the multidimensional surface in $\chi^2$ of the correlated parameters. A 
careful check for ambiguities between the parameters was performed. Besides the well-known correlation between 
$T_{\rm eff}$ and [M/H], we observe in the given temperature range a slight degeneracy between $T_{\rm eff}$ 
and $\log g$ and a strong correlation between $\log g$ and the He content. For the latter, we derive d\,$\log 
g/{\rm d\,He} ={}-$0.15 from our LLmodel atmospheres, where He is the He abundance in units of the solar 
value. An enhanced He abundance generates additional pressure \citep{amar,Fos}, changing in our case $\log g$ 
from 3.36 for solar He abundance to 3.55 for 1.3 times solar abundance.  The effect on $T_{\rm eff}$ is 
marginal, as also found by \citet{Fos}. We determine d\,$T_{\rm eff}{\rm /d\,He} ={}$22 K. No degeneracy 
between any of the other parameters was found. A test of the influence of the abundances of different chemical 
elements showed that only a modification of the He, Ne, Si, S, and Fe abundances had a significant effect on 
the quality of the fit. Except for Ne, the metal abundances turned out to be very nearly solar. The numbers 
are given in the second column of Table 6; the standard deviations were estimated from $\chi^2$ statistics as 
described before.

\begin{figure} 
\includegraphics[width=75mm]{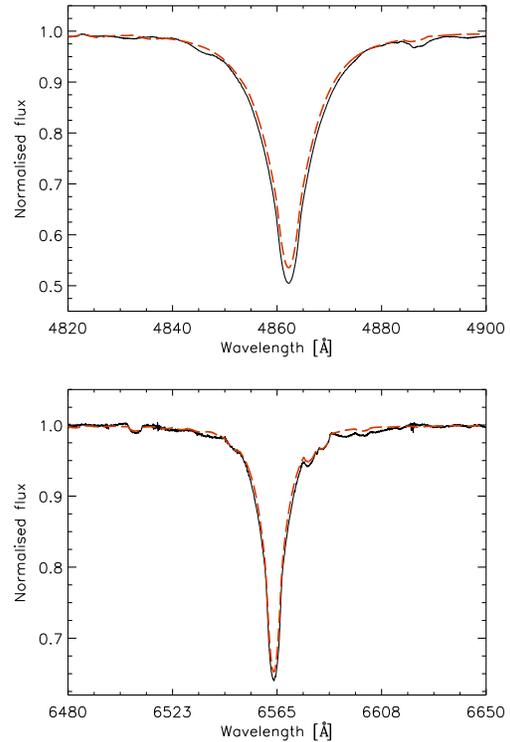} 
\caption{Upper panel: the observed mean profile of the H$\beta$ line (black, solid) and the synthetic 
profile, computed for $T_{\rm eff} ={}$15\,670~K, $\log g ={}$ 3.5 dex, $\xi = 4.0$\,km\,s$^{-1}$, and 
$V_{\rm rot} \sin i ={}$136\,km\,s$^{-1}$ (red, dashed). Lower panel: the same for the H$\alpha$ line. ({\em 
A colour version of this figure is available in the online journal.})}
\label{Fig10}
\end{figure}

EN used the same mean spectrum as HL did, a grid of line-blanketed LTE model atmospheres computed with the 
ATLAS9 code \citep{K93a} and synthetic spectra computed with the SYNTHE code \citep{K93b}. The stellar line 
identification in the entire observed spectral range was done using the line list of \citet{ch}. To begin 
with, the photometric $T_{\rm eff}$ and $\log g$ (see Sect.\,7.1) were verified by comparing the observed 
profiles of the H$\beta$ and H$\alpha$ lines with the synthetic profiles, computed for $T_{\rm eff}$ from 
15\,000 K to 16\,000 K with a step of 50 K, and $\log g$ from 3.0 dex to 4.0 dex with a step of 0.1 dex. The 
best match was obtained for 15500 K${}< T_{\rm eff} <{}$15700 K and 3.5 dex${}< \log g <{}$ 3.6 dex, in 
excellent agreement with the photometric values and with HL. Then, parts of the spectrum  suitable for the 
analysis were selected. Next, the elements which influence every chosen part were identified. Finally, 
assuming the photometric $T_{\rm eff}$ and $\log g$, $\xi = 4.0$\,km\,s$^{-1}$, and solar helium abundance, 
the values of $V_{\rm rot} \sin i$ and the abundances were determined by means of the spectrum synthesis 
method described in \citet{NMA09}. The rotational velocity was derived from every part of the spectrum, 
whereas the abundance of a given element was determined from all parts in which this element was identified. 
The final values of $V_{\rm rot} \sin i$, and of the abundances are average values obtained from all parts in 
which they were determined. The final value of $V_{\rm rot}\sin i$ is equal to 136$\,\pm\,$6 km\,s$^{-1}$. A 
comparison of the TLS mean profiles of the H$\beta$ and H$\alpha$ lines with the synthetic profiles, computed 
with the photometric $T_{\rm eff}$ and $\log g$, $\xi = 4.0$\,km\,s$^{-1}$, and $V_{\rm rot}\sin i ={}$ 
136\,km\,s$^{-1}$, is presented in Fig.~\ref{Fig10}.

The abundances with their standard deviations are given in the third column of Table 6, and the number of 
parts of the spectrum from which they were derived, $n$, in column four. The solar abundances listed in the 
last column are quoted from \citet{agj}. Both, HL and EN determined the abundances of Ne, Si, S and Fe. Except 
for Si, the agreement between the HL and EN abundances is within 1\,$\sigma$. In the case of Si, HL's $\log N$ 
is virtually identical with $\log N_{\sun}$. As can be seen from Table 6, the C, N and O abundances determined 
by EN are slightly smaller then solar, whereas Al (only one part of the spectrum used) and P are enhanced. The 
abundances of Mg, S, Cl, Fe and Ni obtained by EN are close to the solar values. On the other hand, Ne is 
overabundant by 0.71$\,\pm\,$0.11\,dex according to EN, and by 0.56$\,\pm\,$0.15 \,dex according to HL. The Ne 
abundances of B-type stars were investigated by \citet{cuh} and \citet{mb08}. In these papers a non-LTE 
approach was adopted; the Ne abundances  turned out to be equal to 8.11$\,\pm\,$0.04 and 
7.97$\,\pm\,$0.07\,dex, respectively. In addition, \citet{cuh} found that the difference between the LTE and 
non-LTE abundance derived from the Ne\,I 6402 \AA\ line amounted to 0.56\,dex. This result is close to the 
difference between the HL and EN LTE-determinations and the non-LTE values of \citet{cuh} and \citet{mb08}. 

\begin{table}
 \centering
 \begin{minipage}{80mm}
\caption{The elemental abundances of $\mu$\,Eri derived by HL and EN; $n$ is explained in the text.}\vspace{2mm}
\begin{center}
\begin{tabular}{cccrr}
\hline
\multicolumn{1}{c}{Element} & \multicolumn{1}{c}{$\log N$}&  
\multicolumn{1}{c}{$\log N$} &\multicolumn{1}{c}{$n$} & \multicolumn{1}{c}{$\log N_{\sun}$}\\
\multicolumn{1}{c}{} & \multicolumn{1}{c}{HL} & \multicolumn{1}{r}{EN}&\\
\hline
He&11.04$\pm$0.034  &--&  --&  10.93\\
C &--&              8.26$\pm$0.27&  6& 8.39\\
N &--&              7.41$\pm$0.29&  3& 7.78\\
O &--&              8.49$\pm$0.31&  3& 8.66\\
Ne&8.40$\pm$0.26&   8.55$\pm$0.11& 13& 7.84\\
Mg&--&              7.51&           2& 7.53\\
Al&--&              6.77&           1& 6.37\\
Si&7.49$\pm$0.21&   7.11$\pm$0.12&  7& 7.51\\
P &--&              5.82$\pm$0.21&  5& 5.36\\
S &6.94$\pm$0.20&   7.06$\pm$0.18& 18& 7.14\\
Cl&--&              5.51&           1& 5.50\\
Fe&7.45$\pm$0.30&   7.34$\pm$0.23& 20& 7.45\\
Ni&--&              6.21$\pm$0.15&  3& 6.23\\
\hline
\end{tabular}
\end{center}
\end{minipage}
\end{table}

In addition to differences in methodology and software, an important difference between the HL and EN analyses 
is the value of the microturbulence velocity. While HL derived $\xi ={}$0.0\,km\,s$^{-1}$, EN assumed 4.0 
\,km\,s$^{-1}$ following \citet{Lef+10}. The main consequence of this difference is the following: For a 
strong line, greater $\xi$ results in a greater equivalent width of the line in the synthetic spectrum, so 
that matching the synthetic and observed line-profiles requires a smaller abundance than would be the case for 
a smaller $\xi$. The determination of the Si abundance is dominated  by the strong Si\,II doublet at 6347/6371 
{\AA}. This explains why EN obtained appreciably smaller abundance of Si than did HL (see Table 6). In the 
case of Fe, there was one strong line, viz., Fe\,I at 5057 {\AA}, and dozens of fainter lines; consequently, 
the effect is smaller than for Si. The effect is absent in the case of Ne and S because no strong lines were 
used in the analysis. 

\section{Mean density and the HR diagram}

Using the effective temperature and luminosity obtained in Sect.\ 7.1 and assuming a mass of 6.2$\,\pm\,$0.2 
M$_{\sun}$ (see below) we find the mean density of the primary component, $\langle \rho \rangle_{\rm ph} 
={}$0.0418$\,\pm\,$0.0057 g\,cm$^{-3}$, where the suffix ``ph'' is a reminder of the photometric provenance 
of this value. Surprisingly, the largest contribution to the large standard deviation does not come from the 
standard deviation of $T_{\rm eff}$, but from that of $L$: if the latter were omitted, the standard 
deviation of $\langle \rho \rangle_{\rm ph}$ would be reduced to 0.0033 g\,cm$^{-3}$; omitting the former 
yields a standard deviation of 0.0048 g\,cm$^{-3}$. 

A value of the mean density independent of photometric calibrations can be derived in the present case 
because for a binary with the primary's mass much larger than that of the secondary, the spectroscopic and 
photometric elements can be combined to derive a mass-radius relation 
\begin{equation} 
M_{\rm p} \sim R_{\rm p}^{3+x}, 
\end{equation}
where $R_{\rm p}$ is the radius of the primary and $x$ is a small positive number, slowly varying with $M_{\rm 
p}$, the assumed mass of the primary \citep{pj}. A consequence of this relation is that the mean density of 
the primary computed from the orbital solution, $\langle \rho \rangle_{\rm orb}$, is weakly dependent on 
$M_{\rm p}$. This fact has been used by \cite{dj} to derive a mean density of the $\beta$ Cephei primary of 16 
Lac (EN Lac), an SB1 and EA system rather similar to $\mu$ Eri. In the present case, the spectroscopic 
parameters $K_1$ and $e$ from Table 3 and the photo\-metric parameters $r_{\rm p}$ and $i$ from line (3) or 
(4) of Table 4 give $\langle \rho \rangle_{\rm orb} ={}$0.0345 g\,cm$^{-3}$, 0.0348 g\,cm$^{-3}$, and 0.0350 
g\,cm$^{-3}$ for $M_{\rm p} ={}$5 M$_{\sun}$, 6 M$_{\sun}$, and 7 M$_{\sun}$, respectively, with the standard 
deviations equal to 0.0032 g\,cm$^{-3}$. The last value is determined mainly by the standard deviation of 
$K_1$: if the standard deviation of $K_1$ were omitted, the standard deviation of $\langle \rho \rangle_{\rm 
orb}$ would be equal to 0.0012 g\,cm$^{-3}$; the contributions of the standard deviations of the photometric 
para\-meters are negligible. For $M_{\rm p} ={}$6.2~M$_{\sun}$, the difference between the two values of mean 
density, $\Delta \langle \rho \rangle = \langle \rho \rangle_{\rm ph} - \langle \rho \rangle_{\rm orb}$, is 
equal to $0.0070\,\pm\,0.0065$~g\,cm$^{-3}$. If the photometric parameters from line (1) or (2) were used, 
$\langle \rho \rangle_{\rm orb}$ would be 0.0010 g\,cm$^{-3}$ smaller than those given above and $\Delta 
\langle \rho \rangle$ would amount to $0.0080\,\pm\,0.0065$ g\,cm$^{-3}$. In what follows we use the larger 
values of $\langle \rho \rangle_{\rm orb}$, i.e., those obtained with the photometric parameters from line (3) 
or (4) of Table 4. 

\begin{figure} 
\includegraphics[width=80mm]{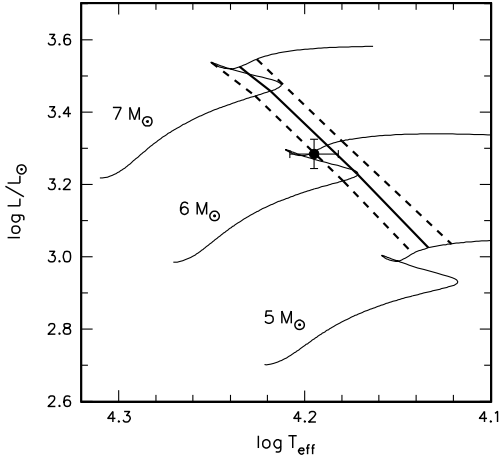} 
\caption{$\mu$ Eri in the HR diagram. The point with error bars was plotted using the photometric effective 
temperature and the luminosity computed from the revised {\em Hipparcos\/} parallax and the photometric BC 
(see Section 7.1). The thin solid lines are $X ={}$0.7, $Z ={}$0.02 evolutionary tracks for 5 M$_{\sun}$, 6 
M$_{\sun}$, and 7~M$_{\sun}$. The thick solid line connects points for which the primary component's mean 
density derived from the orbital solution, $\langle \rho \rangle_{\rm orb}$, is equal to the evolutionary 
tracks' mean density. The dashed lines run at $\langle \rho \rangle_{\rm orb}\,\pm\,$2$\sigma$.} 
\label{Fig11} 
\end{figure}

Figure~\ref{Fig11} shows an HR diagram in which $\mu$ Eri is plotted using the effective temperature and 
luminosity obtained in Section 7.1. Also shown are $X ={}$0.7, $Z ={}$0.02 evolutionary tracks for 5 
M$_{\sun}$, 6 M$_{\sun}$, and 7 M$_{\sun}$. The tracks were kindly computed for us by Dr.\ J.\ 
Daszy\'nska-Daszkiewicz with the OPAL opacities, A04 mixture, no convective-core overshooting, and $V_{\rm 
rot} ={}$160 km\,s$^{-1}$ on the ZAMS. The code was the same as that used by \citet{ddp08}. The position of 
the star in relation to the evolutionary tracks is very nearly the same as that in figure 5 of \citet{hsj}. 
Thus, these authors' conclusion that $\mu$ Eri is just before or shortly after the end of the main-sequence 
phase of evolution (see the Introduction) remains unaltered. As can be seen from Fig.~\ref{Fig11}, the 
assumption that the star is still on the main sequence leads to $M_{\rm p} ={}$6.2$\,\pm\,$0.2~M$_{\sun}$. 

The thick solid line in Fig.~\ref{Fig11} connects points for which $\langle \rho \rangle_{\rm orb}$ is equal 
to the evolutionary tracks' mean density. The dashed lines run at $\langle \rho \rangle_{\rm 
orb}\,\pm\,$2$\sigma$. For $M_{\rm p} \ga{}$5~M$_{\sun}$, the line crosses the tracks at an advanced stage of 
hydrogen burning in the core, a phase of secondary contraction, or a phase of shell hydrogen-burning. This is 
the same conclusion as that of \citet{hsj} but independent of photometric calibrations and the {\em 
Hipparcos\/} parallax.  

\section{The SPB terms}

\subsection{A comparison with the ground-based results}

\begin{figure} 
\includegraphics{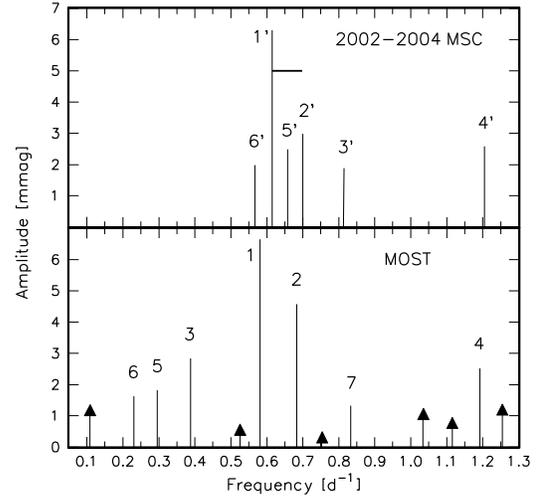} 
\caption{A comparison of the 2002-2003 and 2003-2004 multi-site photometric campaigns (MSC) SPB terms (upper 
panel) with the 13 strongest terms from Table 1 having $f_j <{}$1.3 d$^{-1}$ (lower panel). The MSC $f'_j$ 
($j = 1,..,6$) terms are labeled in the upper panel, and the MOST $f_j$ ($j=1,..,7$) terms, in the lower 
panel; the six fainter MOST terms are marked with solid triangles. The horizontal line beginning at $f'_1$ 
in the upper panel has a length equal to 0.083~d$^{-1}$, the frequency resolution of the MOST data.} 
\label{Fig12} 
\end{figure}

The six SPB terms, derived from the 2002-2003 and 2003-2004 multi-site photometric campaigns (MSC, see the 
Introduction), are compared in Fig.~\ref{Fig12} with the 13 strongest terms from Table 1 having $f_j 
<{}$1.3~d$^{-1}$. In the upper panel are plotted the $v$-filter amplitudes of the $f'_j$ ($j=1,..,6$) 
terms from table 7 of JHS05, i.e., those obtained from the combined 2002-2003 and 2003-2004 MSC data. In the 
lower panel are plotted the MOST amplitudes of the $f_j$ ($j ={}$1,..,11, 13, and 16) terms from Table 
1; the seven strongest ones are labeled, the remaining ones are marked with solid triangles. The horizontal 
line beginning at $f'_1$ in the upper panel has a length equal to the frequency resolution of the MOST 
data. Clearly, the $f'_6,f'_1,f'_5,f'_2$ quadruplet would not be resolved in the frequency analysis in 
Sect.\ 3.1. 

In the time domain, the situation is the following. The $f'_6,f'_1,f'_5,f'_2$ quadruplet gives rise to a 
multiperiodic beat-phenomenon. The shortest beat-period is equal to 7.5 d, the three longest ones range from 
20.9 d to 23.8 d, and the dominant beat-period, due to the interference between the largest-amplitude terms 
$f'_1$ and $f'_2$, is equal to 11.8 d.  If the quadruplet were present in the light variation of $\mu$ Eri 
when the star was observed by MOST, the 12-day light-curve seen in Fig.~\ref{Fig01} would include one 
dominant beat-cycle and only halves of the longest ones. The question now is whether the short time-span of 
the MOST data or, equivalently, its limited frequency resolution, could be the sole reason for the appearance 
of the two frequencies, $f_1$ and $f_2$, in the lower panel of Fig.~\ref{Fig12} instead of the quadruplet.  

In order to answer this question we carried out frequency analysis of two sets of synthetic time-series. 
Both sets consisted of the magnitudes, $m_i$, computed from the equation
\begin{equation} 
m_i = \sum_{j=1}^6 A'_j \sin (2 \pi f'_j t_i + \Phi'_j), 
\end{equation} 
where $t_i$ are the epochs of the MOST observations. In set 1, the amplitudes, $A'_j$, and the frequencies, 
$f'_j$, were taken from table 7 of JHS05, i.e., they were equal to those plotted in the upper panel of 
Fig.~\ref{Fig12}. The initial phases, $\Phi'_j$, were adjusted so that the rms difference between the MOST 
out-of-eclipse detrended light-curve and the synthetic light-curve was a minimum. In this way we have shifted 
the synthetic beat-pattern in time to best represent the observed variation. Note that because of the lapse of 
time, the MSC initial phases are useless in this context. 

In set 2, only the frequencies, $f'_j$, were from table 7 of JHS05, while the amplitudes, $A'_j$, and 
the initial phases, $\Phi'_j$, were adjusted to minimize the rms difference between the MOST and synthetic 
light-curves. In other words, the two sets represent error-free 2002-2004 light-curves of $\mu$ Eri sampled 
at the epochs of the MOST observations, with adjusted initial phases in set 1, and the amplitudes and phases 
in set~2. 

\begin{figure} 
\includegraphics{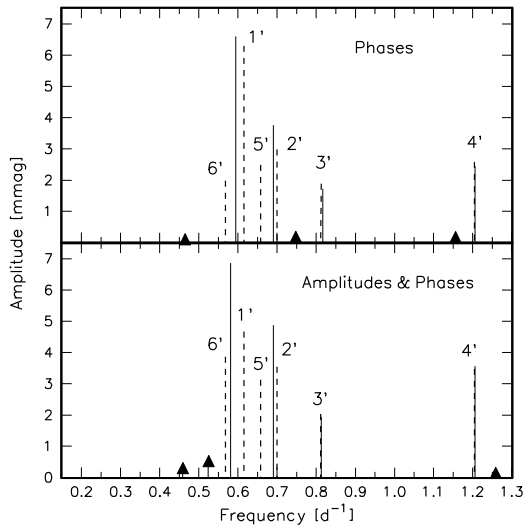} 
\caption {The results of a frequency analysis of synthetic time-series having the same frequency resolution 
as the MOST data (solid lines and triangles). The MSC $f'_j$ ($j = 1,..,6$) terms (dashed lines) are shown 
for comparison. Upper panel: the synthetic time-series was computed with the MSC amplitudes and frequencies 
but the initial phases adjusted to minimize the rms difference between the MOST light-curve and the synthetic 
one (referred to as set 1 in the text). Lower panel: the same for set 2 in which not only the initial phases 
but also the amplitudes were adjusted. In the upper panel the dashed lines show the MSC amplitudes (the same 
as the solid lines in Fig.~\ref{Fig12}), while in the lower panel are shown the adjusted amplitudes.} 
\label{Fig13} 
\end{figure}

The results of a frequency analysis of sets 1 and 2, carried out in the same way as the frequency analysis 
in Sect.\ 3.1, are shown in the upper and lower panel of Fig.~\ref{Fig13}, respectively. In both cases, the 
seven strongest terms are displayed. As can be seen from the figure, only four terms (solid lines) have 
survived with sizable amplitude. The remaining three terms (triangles) have amplitudes ranging from 0.192 
mmag to 0.105 mmag in the upper panel, and from 0.524 mmag to 0.144 mmag in the lower panel. In both panels, 
the $f'_6,f'_1,f'_5,f'_2$ quadruplet is replaced by two terms with the amplitudes and frequencies rather 
close to those of the MOST $f_1$ and $f_2$ terms. In the upper panel, the first term (the highest-amplitude 
solid line) has an amplitude very nearly equal to that of the $f_1$ term and the frequency 0.013~d$^{-1}$ 
greater than $f_1$. The second term (the second highest-amplitude solid line) has an amplitude 0.81 mmag 
smaller than that of the $f_2$ term and the frequency 0.007 d$^{-1}$ greater than $f_2$. In the lower panel, 
the first term matches the $f_1$ term almost perfectly: it has the same frequency and  amplitude only 
0.26 mmag greater. The second term fares only slightly less well: its frequency and amplitude differ from 
those of the $f_2$ term by 0.008 d$^{-1}$ and 0.32 mmag, respectively. Thus, the question posed at the 
beginning of this section can be answered in the affirmative: the limited frequency resolution of the MOST 
data is the sole reason for the appearance of the two frequencies, $f_1$ and $f_2$, instead of the 
$f'_6,f'_1,f'_5,f'_2$ quadruplet. We conclude that the four frequencies have not changed appreciably during 
the six years between the epochs of the MSC and MOST observations. In addition, the dashed lines in the two 
panels of Fig.~\ref{Fig13} show no amplitude change of $f'_5$ and $f'_2$, but a decrease in the amplitude of 
$f'_1$, and an increase in that of $f'_6$, indicating long-term amplitude variability of these two terms of 
the quadruplet. 

>From Fig.~\ref{Fig13}, one can also see that terms $f'_3$ and $f'_4$ are retrieved almost unchanged. The 
MOST terms closest in frequency to $f'_3$ and $f'_4$ are $f_7$ and $f_4$, respectively. As can be seen from  
Fig.~\ref{Fig12}, terms $f'_4$ and $f_4$ are close to each other in amplitude and frequency. We conclude 
that they are almost certainly due to the same underlying variation, unchanged since 2002-2004. The case of 
$f'_3$ and $f_7$ is less certain because the frequency difference amounts now to 0.022~d$^{-1}$, almost 
twice the difference between $f_4$ and $f'_4$. However, given the limited frequency resolution of the MOST 
data, the possibility that $f'_3$ and $f_7$ are related cannot be excluded. 

Let us now consider the small-amplitude synthetic terms, shown with triangles in Fig.~\ref{Fig13}. The 
highest-amplitude term, at 0.525~d$^{-1}$ in the lower panel of the figure, is virtually identical with term 
$f_{13}$ of Table 1. Another one, at 0.748~d$^{-1}$ in the upper panel, is close in frequency to term 
$f_{16}$ but has an amplitude smaller by a factor of about 0.4. The remaining ones bear no relation to the 
terms listed in Table 1. 

\subsection{SPB terms added by MOST}

The low-frequency terms $f_3$, $f_5$, $f_6$ (solid lines), and $f_9$ (the triangle at 0.109 d$^{-1}$), 
plotted in the bottom panel of Fig.~\ref{Fig12}, have no counterparts in the MSC frequency spectrum of $\mu$ 
Eri. This does not necessarily mean that these terms were absent in the star's light-variation in 2002-2004. 
They all have amplitudes smaller than 2.9 mmag, and therefore may have been missed by JHS05 because of the 
$1/f$ increase of periodogram noise in ground-based photometry. 

The three terms seen close to $f_4$ in the lower panel of Fig.~\ref{Fig12}, viz., $f_{10}$ at 
1.035~d$^{-1}$, $f_{11}$ at 1.115~d$^{-1}$, and $f_8$ at 1.254~d$^{-1}$, have amplitudes below the detection 
threshold of MSC. With $f_4$, they form a quadruplet consisting of a barely-resolved equidistant triplet 
$f_{10}$, $f_{11}$, $f_4$, and the rightmost term $f_8$, separated from $f_4$ by about 3/4 of the frequency 
resolution of the data. 

As can be seen from Table 1, all terms with frequencies higher than 1.3~d$^{-1}$ have amplitudes smaller 
than 0.71 mmag. The two frequencies of largest amplitude are $f_{12}$ at 1.825~d$^{-1}$ and $f_{14}$ at 
2.423~d$^{-1}$; $f_{12}$, $f_{15}$ at 1.620~d$^{-1}$, and $f_{19}$ at 1.722~d$^{-1}$ form an equidistant 
triplet with a separation of 0.103~d$^{-1}$, while $f_{14}$ is preceded by terms $f_{18}$ and $f_{24}$, and 
followed by $f_{17}$ and $f_{20}$. The frequency separation between the preceding terms is very nearly equal 
to that between the following terms, and is equal to the separation between the members of the $f_{12}$, 
$f_{15}$, $f_{19}$ triplet. 

The remaining terms with frequencies lower than 15~d$^{-1}$, viz., $f_{22}$ at 3.780~d$^{-1}$, $f_{27}$ at 
4.850~d$^{-1}$, and $f_{31}$ at 5.499~d$^{-1}$, have $S/N \le{}$3.5, and therefore are insignificant 
\citep{Bre93}. Thus, the lowest SPB frequency in the light-variation of $\mu$ Eri we have found in the MOST 
data is $f_9 ={}$0.109~d$^{-1}$, while the highest one is $f_{20} ={}$2.786~d$^{-1}$. However, keeping in 
mind the synthetic time-series lesson of the preceding section, we warn that neither these frequencies nor 
the frequencies in the multiplets around $f_4$, $f_{12}$, and $f_{14}$ should be taken at their face values. 

\section{Summary and suggestions for improvements}

A frequency analysis of the MOST time-series photometry of $\mu$ Eri (Sect.\ 2) resulted in decomposing the 
star's light-variation into a number of sinusoidal terms (Table 1) and the eclipse light-curve 
(Fig.~\ref{Fig05}). New radial-velocity observations (Sect.\ 4.1), some obtained contemporaneously with the 
MOST photometry, were used to compute parameters of a spectroscopic orbit, including the semi-amplitude of 
the primary's radial-velocity variation, $K_1$, and the eccentricity, $e$ (Table 3). Frequency analysis of 
the radial-velocity residuals from the spectroscopic orbital solution failed to uncover intrinsic variations 
with amplitudes greater than 2 km\,s$^{-1}$ (Sect.\ 4.3). The eclipse light-curve was solved by means of the 
computer program {\sc EBOP} (Sect.\ 5). The solution confirms that the eclipse is a total transit, yielding 
the inclination of the orbit, $i$, and the relative radius of the primary, $r_{\rm p}$ (Table 4). A 
Rossiter-McLaughlin anomaly, implying the primary's direction of rotation concordant with the direction of 
orbital motion and $V_{\rm rot} \sin i$ of 170$\,\pm\,$45 km\,s$^{-1}$, was detected in observations covering 
ingress (Sect.\ 6). 

>From archival $UBV$ and $uvby\beta$ indices and the revised {\em Hipparcos\/} parallax, we derived 
the primary's effective temperature, $T_{\rm eff} ={}$15\,670$\,\pm\,$470 K, surface gravity, $\log g 
={}$3.50$\,\pm\,$0.15, bolometric correction, BC $={}-$1.40$\,\pm\,$0.08 mag, and the luminosity, $\log 
L/{\rm L}_{\sun} ={}$3.280$\,\pm\,$0.040 (Sect.\ 7.1). These values of $T_{\rm eff}$ and $\log L/{\rm 
L}_{\sun}$ place the star in the HR diagram (Fig.~\ref{Fig11}) in very nearly the same position as in figure 
5 of \citet{hsj}. Thus, these author's conclusion that $\mu$ Eri is just before or shortly after the end of the 
main-sequence stage of evolution remains unaltered.  A discrepancy between the effective temperature and 
bolometric correction we derived and those obtained from \citet*{f96} calibration of the $B-V$ index was noted. 
An analysis of a high signal-to-noise ratio spectrogram yielded $T_{\rm eff}$ and $\log g$ in good 
agreement with the photometric values. The abundance of He, C, N, O, Ne, Mg, Al, Si, P, S, Cl, and Fe were 
obtained from the same spectrum by means of the spectrum synthesis method. 

A mean density, $\langle \rho \rangle_{\rm ph} ={}$0.0418$\,\pm\,$0.0057 g\,cm$^{-3}$, was computed from the 
star's HR coordinates and the mass $M_{\rm p} ={}$6.2$\,\pm\,$0.2 M$_{\sun}$, estimated from an $X ={}$0.7, $Z 
={}$0.02 evolutionary track under assumption that the star is still in the main-sequence stage of its 
evolution. Another value of the mean density, $\langle \rho \rangle_{\rm orb} \approx{}$0.035$\,\pm\,$0.003 
g\,cm$^{-3}$, very nearly independent of $M_{\rm p}$, was computed from $K_1$, $e$, $i$ and $r_{\rm p}$ 
(Sect.~8). With the help of evolutionary tracks, this allowed placing the star in the HR diagram on a line 
$\langle \rho \rangle_{\rm orb} \approx const$. For evolutionary tracks with $X ={}$0.7, $Z ={}$0.02, this 
line runs close to the star's position mentioned above. For $M_{\rm p} \ga{}$5~M$_{\sun}$, the line crosses 
the tracks at an advanced stage of hydrogen burning in the core, a phase of secondary contraction, or a phase 
of shell hydrogen-burning (Fig.~\ref{Fig11}). This is the same conclusion as that of \citet{hsj} but 
independent of photometric calibrations and the {\em Hipparcos\/} parallax. 

The difference between the two values of mean density of the primary component of $\mu$ Eri, $\Delta \langle 
\rho \rangle = \langle \rho \rangle_{\rm ph} - \langle \rho \rangle_{\rm orb}$, is equal to 
$0.0070\,\pm\,0.0065$ g\,cm$^{-3}$ (Sect 8). The difference, although large, is not significant. Thus, the 
discrepancy mentioned in the Introduction has vanished. It is now clear that the value of $\langle \rho 
\rangle_{\rm orb}$ derived by \citet{j09}, viz., 0.0132$\,\pm\,$0.0050 g\,cm$^{-3}$, was too small by a 
factor of 2.6. The sole reason for this is that the relative radius of the primary component obtained from 
the ground-based photometry, $r_{\rm p} ={}$0.287, is 1.4 times greater than $r_{\rm p}$ from the MOST 
photometry (Table 4). This, in turn, was caused by insufficient accuracy of the ground-based photometry. 

The standard deviation of $\Delta \langle \rho \rangle$ cannot be easily reduced because it is determined 
mainly by the standard deviation of $\langle \rho \rangle_{\rm ph}$ which, in turn, is dominated by the 
standard deviation of the luminosity (Sect.\ 8). Reducing the latter would require more accurate bolometric 
corrections (Sect.\ 7.1) or, in other words, better absolute bolometric fluxes of hot stars than those 
provided years ago by \citet{cod}. As far as we are aware, such data are not forthcoming. On the other hand, 
$\langle \rho \rangle_{\rm orb}$ can be improved readily by obtaining a more accurate value of $K_1$ than that 
given in Table 3. A radial-velocity project undertaken to achieve this purpose should include measurements 
covering the eclipse so that the velocity of rotation of the primary component could be better constrained 
than in Sect.\ 6. 

The frequency resolution of the MOST observations of $\mu$ Eri is insufficient to resolve the closely-spaced 
SPB frequency quadruplet, 0.568 d$^{-1}$, 0.616 d$^{-1}$, 0.659 d$^{-1}$, and 0.701 d$^{-1}$, discovered 
from the ground by JHS05. Despite this, we showed that the four frequencies have not changed appreciably 
during the six years between the epochs of the ground-based and MOST observations (Sect.\ 9.1). The other 
two SPB frequencies seen from the ground, 0.813~d$^{-1}$ and 1.206 d$^{-1}$, may also have their MOST 
counterparts. Moreover, our analysis of the MOST data added 15 SPB terms with frequencies ranging from 
0.109 d$^{-1}$ to 2.786~d$^{-1}$; the amplitudes range from 2.843$\,\pm\,$0.013 mmag at 0.387 d$^{-1}$ to 
0.164$\,\pm\,$0.008 mmag at 2.786~d$^{-1}$ (Sect.\ 9.2). However, in view of the limited frequency 
resolution of the MOST data, these frequencies should be treated with caution. Finally, short-period 
variations of uncertain origin were found to be present in the data (the Appendix). 

\section*{Acknowledgments}

We are indebted to Dr.\ Paul B.\ Etzel for providing the source code of his computer program {\sc EBOP} and 
explanations and to Dr.\ Jadwiga Daszy\'nska-Daszkiewicz for computing the evolutionary tracks used in Sect.\ 
8. MJ, EN, and JM-\.Z acknowledge support from MNiSW grant N~N203 405139. EN acknowledges support from NCN 
grant 2011/01/B/St9/05448. Calculations have been partially carried out at the Wroc{\l}aw Centre for Networking 
and Supercomputing under grant No.\ 214. MHr acknowledges support from DFG grant HA 3279/5-1. WD thanks his 
students, Karolina B\c{a}kowska, Adrian Kruszewski, Krystian Kurzawa, and Anna Przybyszewska for assisting in 
observations. DBG, JMM, AFJM and SMR acknowledge funding support of the Natural Sciences and Engineering 
Research Council (NSERC) of Canada. AFJM is also grateful for financial assistance from Fonds de recherche du 
Qu\'ebec (FQNRT). In this research, we have used the Aladin service, operated at CDS, Strasbourg, France, and 
the SAO/NASA Astrophysics Data System Abstract Service.

\appendix
\section{Short-term variations in the MOST observations of $\bmu$ Eri}

\begin{figure} 
\includegraphics{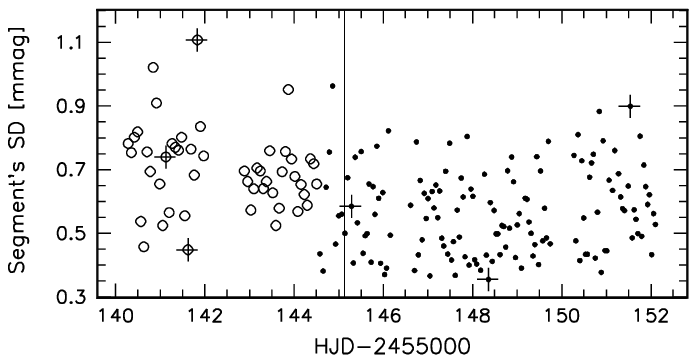} 
\caption {The standard deviations of the residuals from a least-squares fit of equation (2) with $N ={}$20 
and the parameters from Table 1 to the out-of-eclipse data, computed in the data segments and the 0.0433-d 
adjacent intervals of the non-segmented data (the left-hand and the right-hand side of the vertical line, 
respectively). The circles denote segments with $\sim$111 data-points per hour, and the dots, the segments 
and the intervals with $\sim$55 data-points per hour. The pluses indicate the segments and the intervals 
shown in Fig.~\ref{FigA2}.} 
\label{FigA1} 
\end{figure}

As we mentioned in Sect.\ 2, in many data segments and in the non-segmented data there occur short-term 
variations with ranges up to 5 mmag. We shall discuss these variations in some detail presently using the 
residuals, $O-C$, from a least-squares fit of equation (2) with $N ={}$20 and the parameters from Table 1 to 
the out-of-eclipse data. This is justified because all terms up to $f_{20}$ have periods longer than 0.36~d, 
about 8 times the length of the longest segment. In order to include the non-segmented data, we cut them 
into adjacent intervals of constant length, equal to 0.0433 d, the mean length of the segments. Then, we 
computed the standard deviations of the residuals in the segments and the intervals. The results are plotted 
in Fig.~\ref{FigA1}. 

\begin{figure} 
\includegraphics{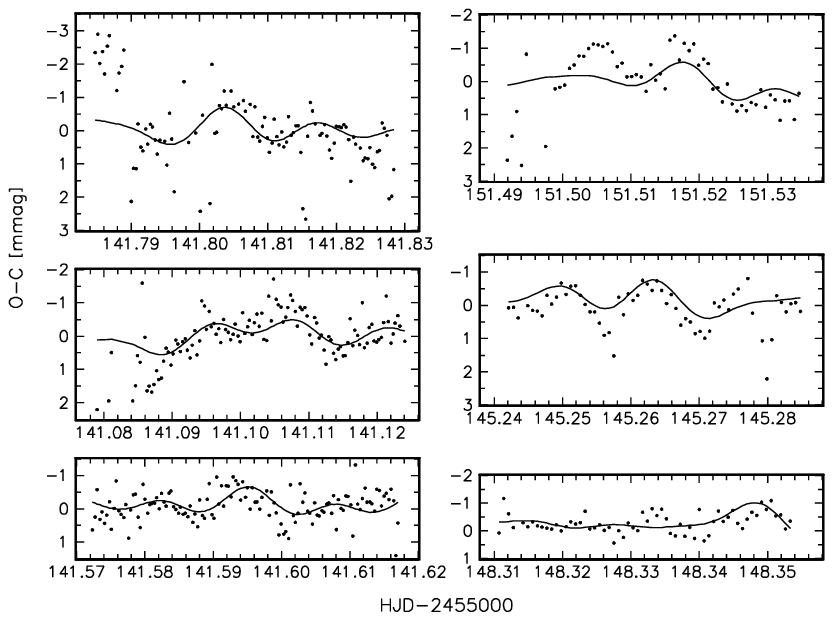} 
\caption {The $O-C$ residuals from  the $N ={}$20 fit for the three segments (the left-hand panels) and 
three 0.0433-d intervals (the right-hand panels) indicated with pluses in Fig.~\ref{FigA1}. Plotted are the 
data (dots) and the $N ={}$31 synthetic light-curve (lines).} 
\label{FigA2} 
\end{figure}

Fig.~\ref{FigA2} shows the segments and the 0.0433-d intervals indicated with pluses in Fig.~\ref{FigA1}. We 
selected the segments and the intervals having the largest, median, and smallest standard deviations (from 
top to bottom, respectively). 

In Sect.\ 3.1, we mentioned that formally significant terms (i.e., those having $S/N >{}$4.0) with $f_j 
>{}$15~d$^{-1}$ have frequencies which are either close to whole multiples of the orbital period of the 
satellite or differ from the whole multiples by $\sim$1~d$^{-1}$. Indeed, denoting the orbital period of the 
satellite by $f_o$ we have $f_{21} \approx{}$5$f_o+$1~d$^{-1}$, $f_{23} \approx{}$2$f_o$, $f_{25} \approx 
f_o+$1~d$^{-1}$, $f_{26} \approx{}$4$f_o+$1~d$^{-1}$, $f_{28} \approx{}$3$f_o$, $f_{29} 
\approx{}$4$f_o-$1~d$^{-1}$, and $f_{30} \approx{}$6$f_o+$1~d$^{-1}$. Although it is not clear why a 
frequency of $\sim$1~d$^{-1}$ should appear in the satellite data, all these frequencies are probably 
instrumental. The remaining $j >{}$20 terms have $S/N <{}$4.0, but were included in order to remove as much 
out-of-the-eclipse variation as possible from the light-curve. From Fig.~\ref{FigA2} it is clear, however, 
that the $N ={}$31 solution leaves a part of the short-period variations unaccounted for. Adding further 
periodic terms did not help. Thus, the unaccounted-for part must consist of incoherent short-period 
variations. Whether they are intrinsic or instrumental is impossible to say. 

\bsp

\label{lastpage}

\end{document}